\documentclass[aps,prc,twocolumn,showpacs,preprintnumbers,
nofootinbib,float,superscriptaddress,longbibliography]{revtex4-1}
\usepackage{graphicx, fancybox}
\usepackage{amsmath,amssymb}
\usepackage[colorlinks=true, pdfstartview=FitV, linkcolor=red, citecolor=blue, urlcolor=blue]{hyperref}
\usepackage[dvipsnames]{xcolor}
\usepackage{soul}
\usepackage{array}
\usepackage{url}
\usepackage[utf8]{inputenc}

\graphicspath{{./}}   
\usepackage[normalem]{ulem}

\begin{document}

\title{Probing early-time longitudinal dynamics with the $\Lambda$ hyperon's spin polarization in relativistic heavy-ion collisions}

\author{Sangwook Ryu}
\email{sangwook.ryu@wayne.edu}
\affiliation{Department of Physics and Astronomy, Wayne State University, Detroit, Michigan, 48201, USA}

\author{Vahidin Jupic}
\email{ep9861@wayne.edu}
\affiliation{Department of Physics and Astronomy, Wayne State University, Detroit, Michigan, 48201, USA}

\author{Chun Shen}
\email{chunshen@wayne.edu}
\affiliation{Department of Physics and Astronomy, Wayne State University, Detroit, Michigan, 48201, USA}
\affiliation{RIKEN BNL Research Center, Brookhaven National Laboratory, Upton, NY 11973, USA}

\begin{abstract}
We systematically study the hyperon global polarization's sensitivity to the collision systems' initial longitudinal flow velocity in hydrodynamic simulations. By explicitly imposing local energy-momentum conservation when mapping the initial collision geometry to macroscopic hydrodynamic fields, we study the evolution of systems' orbital angular momentum (OAM) and fluid vorticity. We find that a simultaneous description of the $\Lambda$ hyperons' global polarization and the slope of pion's directed flow can strongly constrain the size of longitudinal flow at the beginning of hydrodynamic evolution. We extract the size of the initial longitudinal flow and the fraction of orbital angular momentum in the produced QGP fluid as a function of collision energy with the STAR measurements in the RHIC Beam Energy Scan program. We find that there is about 100-200 $\hbar$ OAM that remains in the mid-rapidity fluid at the beginning of hydrodynamic evolution. We further exam the effects of different hydrodynamic gradients on the spin polarization of $\Lambda$ and $\bar{\Lambda}$. The gradients of $\mu_B/T$ can change the ordering between $\Lambda$'s and $\bar{\Lambda}$'s polarization.
\end{abstract}

{\maketitle}

\section{Introduction}

Non-central heavy-ion collisions carry angular momenta of the order of $10^3-10^4 \hbar$. After the initial impact, although most of the angular momentum is carried away by the spectator nucleons, a sizable fraction remains in the created Quark-Gluon Plasma (QGP) and implies a nonzero rotational motion in the fluid. Such rotation inertia can lead to a strong vortical structure inside the resulting liquid. Local fluid vorticity can potentially induce a preferential orientation on the spins of the emitted particles through spin-orbit coupling. The STAR Collaboration at the Relativistic Heavy-Ion Collider (RHIC) discovered the global polarization of $\Lambda$ hyperons, which indicated fluid vorticity of $\omega \approx (9 \pm 1) \times 10^{21} s^{-1}$ \cite{STAR:2017ckg}. This result far surpasses the vorticity of all other known fluids in nature. 
The discovery of global hyperon polarization and 3D simulations of the collision dynamics have opened an entirely new direction of research in heavy-ion physics. To understand the origin of the RHIC $\Lambda$ polarization measurements, we need to address two key theoretical questions: (i) how do the global collision geometry and its orbital angular momentum (OAM) induce the local flow vorticity in heavy-ion collisions? (ii) how do fluid gradients act as thermodynamic forces to polarize the spins of particles?
Resolving these two outstanding questions can provide crucial insights into emergent many-body phenomena in Quantum Chromodynamics (QCD).

Extensive theoretical and phenomenological investigations have been devoted to the effects of fluid vorticity on spin polarization \cite{Liang:2004ph, Becattini:2013fla, Becattini:2016gvu, Karpenko:2016jyx, Xie:2017upb, Karpenko:2021wdm, Huang:2020xyr, Becattini:2020ngo, Huang:2020dtn, Becattini:2020sww, Lisa:2021zkj, Serenone:2021zef, Becattini:2021lfq} as well as the related transport phenomenon involving spin \cite{Jiang:2016woz, Florkowski:2017ruc, Hattori:2019lfp, Liu:2019krs, Fukushima:2020ucl, Liu:2020ymh, Gao:2020vbh, Shi:2020htn, Li:2020eon, Singh:2020rht}. Hydrodynamics + hadronic transport hybrid models and pure transport approaches can provide good descriptions of the global polarization for $\Lambda$ and $\bar{\Lambda}$. 
However, the measured azimuthal distributions of polarization showed an opposite oscillation pattern compared to most of the theoretical results \cite{Becattini:2017gcx, Xia:2018tes, Florkowski:2019voj, Wu:2019eyi, Becattini:2019ntv}.

Most of the phenomenological studies assumed the $\Lambda$'s polarization is directly related to the local thermal vorticity. Recent works \cite{PhysRevLett.94.236601, Mal_shukov_2005, Hidaka:2017auj, Liu:2020dxg, Liu:2021uhn, Becattini:2021suc} proposed that the velocity shear tensor and gradients of $\mu_B/T$ can contribute to the spin polarization of $\Lambda$ and $\bar{\Lambda}$. The effects of velocity shear tensor on the longitudinal polarization's azimuthal dependence were studied and found to be substantial \cite{Fu:2021pok, Becattini:2021iol, Yi:2021ryh}. These results suggest that the hyperon's polarization along the global orbital angular momentum direction is a cleaner observable to study the fluid vorticity evolution in heavy-ion collisions than the measurements of the longitudinal polarization.

This paper will focus on the global $\Lambda$ polarization and study how the measurements can set constraints on the early-time longitudinal dynamics at the RHIC BES energies. In Sec.~\ref{sec:model}, we will introduce a new parametric 3D initial condition model, generalized based on Ref.~\cite{Shen:2020jwv}. In particular, we introduce a model parameter to vary the early-time longitudinal distribution of fluid vorticity. We explicitly impose conservation of orbital angular momentum when mapping the initial collision geometry to hydrodynamic fields. Employing such a model enables us to quantitatively investigate how the global polarization measurements can set constraints on the early-time longitudinal dynamics in heavy-ion collisions. The sensitivity of initial longitudinal flow in pion's directed flow is studied with the same model. In Sec.~\ref{sec:polarization_results}, our phenomenological study will show that a simultaneous description of $\Lambda$ global polarization and the slope of pion's directed flow set strong constraints on the initial condition parameter. The effects of different hydrodynamic gradients on $\Lambda$ polarization will be quantified at the RHIC BES energies.
We will conclude with some closing remarks in Sec.~\ref{sec:conc}.

In this paper we use the conventions for the metric tensor $g^{\mu\nu} = \mathrm{diag}(1, -1, -1, -1)$ and the Levi-Civita symbol $\epsilon^{0123} = 1$.

\section{The Theoretical Framework}
\label{sec:model}

\subsection{Initial-state orbital angular momentum (OAM) and mapping to hydrodynamic fields}

The space-time structure of the initial collision dynamics can be modeled by the 3D MC-Glauber model \cite{Shen:2017bsr, Shen:2020jwv}. We can compute the system's total angular momentum based on the collision geometry before and after the collision impact. Individual nucleon $i$ has its position and momentum $\{x_i^\mu, p_i^\mu\}$. We can compute the relativistic angular momentum as a bivector,
\begin{equation}
    L^{\alpha \beta}_\mathrm{init} \equiv x^\alpha p^\beta - x^\beta p^\alpha,
\end{equation}
which has six independent components.

In fluid dynamics, we can define the angular momentum density tensor,
\begin{equation}
    J^{\mu, \alpha\beta} = x^\alpha T^{\mu \beta} - x^\beta T^{\mu \alpha} + S^{\mu, \alpha \beta}.
\end{equation}
Here the total angular momentum is composed by orbital and spin angular momentum tensors. We can write the orbital angular momentum tensor as
\begin{equation}
    L^{\mu, \alpha\beta} = x^\alpha T^{\mu \beta} - x^\beta T^{\mu \alpha}.
\end{equation}
According to \cite{Misner:1974qy}, we can compute the system's angular momentum tensor on a hyper-surface as,
\begin{equation}
    L^{\alpha \beta}_\mathrm{fluid} = \int d^3 \sigma_\mu L^{\mu, \alpha\beta}.
    \label{eq:OAM_total_surface}
\end{equation}
We choose the hyper-surface along the constant longitudinal proper time $\tau = \sqrt{t^2 - z^2}$,
\begin{equation}
    L^{\alpha \beta}_\mathrm{fluid}(\tau) = \int \tau dx dy d\eta_s L^{\tau, \alpha\beta}.
\end{equation}

In this work, we will exactly match the local energy and momentum from initial collision geometry to the hydrodynamic fields at hydrodynamic starting time $\tau = \tau_0$.
This matching is done at each point on the transverse plane, so that it ensures the system's OAM is preserved from the initial state to the hydrodynamic phase,
\begin{equation}
    L^{\alpha \beta}_\mathrm{init}= L^{\alpha \beta}_\mathrm{fluid}(\tau_0).
\end{equation}
We generalize the geometric-based 3D initial conditions in Ref.~\cite{Shen:2020jwv}. Based on the Glauber geometry, the area density of energy and longitudinal momentum at a given transverse position is given by,
\begin{eqnarray}
    \frac{d}{d^2 \textbf{x}_T} E(x, y) &=& [T_A(x, y) + T_B(x, y)] m_N \cosh(y_\mathrm{beam}) \nonumber \\
    &\equiv& M(x, y) \cosh(y_\mathrm{CM})\\
    \frac{d}{d^2 \textbf{x}_T} P_z(x, y) &=& [T_A(x, y) - T_B(x, y)] m_N \sinh(y_\mathrm{beam}) \nonumber \\
    &\equiv& M(x, y) \sinh(y_\mathrm{CM}).
\end{eqnarray}
Here $T_{A(B)}(x, y)$ is the participant thickness function in the tranvserse plane, $m_N$ is the mass of the nucleon, and $y_\mathrm{beam} \equiv \mathrm{arccosh}[\sqrt{s_\mathrm{NN}}/(2 m_N)]$ is the beam rapidity. 
We define the colliding nucleus $A$ as the projectile with positive rapidity, while the nucleus $B$ is the target flying toward the $-z$ direction. The invariant mass and center-of-mass rapidity can be expressed in terms of the participant thickness functions as follows,
\begin{eqnarray}
    M(x, y) &=& m_N \sqrt{T_A^2 + T_B^2 + 2 T_A T_B \cosh(2 y_\mathrm{beam})} \\
    y_\mathrm{CM}(x, y) &=& \mathrm{arctanh} \left[ \frac{T_A - T_B}{T_A + T_B} \tanh(y_\mathrm{beam}) \right].
\end{eqnarray}
Requiring the energy and momentum to be conserved when mapping the initial condition to hydrodynamic fields, we get the following constraints on the system's energy-momentum tensor,
\begin{eqnarray}
    && M(x, y) \cosh[y_\mathrm{CM}(x, y)] = \int \tau_0 d \eta_s [T^{\tau \tau}(x, y, \eta_s) \cosh(\eta_s) \nonumber \\
    && \qquad \qquad \qquad \qquad \qquad + \tau_0 T^{\tau \eta}(x, y, \eta_s) \sinh(\eta_s)] \label{eq:rule1} \\
    && M(x, y) \sinh[y_\mathrm{CM}(x, y)] = \int \tau_0 d \eta_s [T^{\tau \tau}(x, y, \eta_s) \sinh(\eta_s) \nonumber \\
    && \qquad \qquad \qquad \qquad \qquad + \tau_0 T^{\tau \eta}(x, y, \eta_s) \cosh(\eta_s)] \label{eq:rule2}.
\end{eqnarray}
Here $T^{\tau \tau}(x, y, \eta_s)$ and $T^{\tau \eta}(x, y, \eta_s)$ are components of the system's energy-momentum tensor on a constant proper time hyper-surface with $\tau = \tau_0$.
We assume the initial energy-momentum current has the following form,
\begin{eqnarray}
    T^{\tau\tau} (x, y, \eta_s) &=& e(x, y, \eta_s) \cosh(y_L) \label{eq:initial_Ttautau} \label{eq:Ttautau} \\
    T^{\tau \eta} (x, y, \eta_s) &=& \frac{1}{\tau_0} e(x, y, \eta_s) \sinh(y_L). \label{eq:Ttaueta}
    \label{eq:initial_Ttaueta}
\end{eqnarray}
We ignore the transverse expansion and set transverse components $T^{\tau x} = T^{\tau y} = 0$ at $\tau = \tau_0$.
Here we parameterize a non-zero longitudinal momentum with the rapidity variable
\begin{equation}
    y_L \equiv f y_\mathrm{CM}, \label{eq:yL_def}
\end{equation}
where $f \in [0, 1]$ is a parameter that controls the fraction of longitudinal momentum attributed to the flow velocity. When $f = 0$, $y_L = 0$, the conditions reduce to the well-known Bjorken flow scenario, which was used in Ref.~\cite{Shen:2020jwv}. This longitudinal momentum fraction parameter $f$ allows us to vary the size of the initial longitudinal flow while keeping the net longitudinal momentum of the hydrodynamic fields fixed. 
Plugging Eqs.~(\ref{eq:initial_Ttautau}) and (\ref{eq:initial_Ttaueta}) into Eqs.~(\ref{eq:rule1}) and (\ref{eq:rule2}), we get
\begin{eqnarray}
    M(x, y) &=& \int \tau_0 d \eta_s e(x, y, \eta_s) \cosh(y_L + \eta_s - y_\mathrm{CM}) \\
    0 &=& \int \tau_0 d \eta_s e(x, y, \eta_s) \sinh(y_L + \eta_s - y_\mathrm{CM})). \quad
\end{eqnarray}
To satisfy these two equations, we can choose a symmetric rapidity profile parameterization w.r.t $y_\mathrm{CM} - y_L$ for the local energy density \cite{Hirano:2005xf},
\begin{eqnarray}
    && e (x, y, \eta_s; y_\mathrm{CM} - y_L) = \nonumber \\
    && \qquad \mathcal{N}_e(x, y) \exp\bigg[- \frac{(\vert \eta_s - (y_\mathrm{CM} - y_L) \vert  - \eta_0)^2}{2\sigma_\eta^2} \nonumber \\
    && \qquad \qquad \qquad \qquad \times \theta(\vert \eta_s - (y_\mathrm{CM} - y_L) \vert - \eta_0)\bigg].
    \label{eq:eprof}
\end{eqnarray}
Here the parameter $\eta_0$ determines the width of the plateau and the $\sigma_\eta$ controls how fast the energy density falls off at the edge of the plateau. In a highly asymmetric situation $T_A(x, y) \gg T_B(x, y)$, the center-of-mass rapidity $y_\mathrm{CM}(x, y) \rightarrow y_\mathrm{beam}$. To make sure there is not too much energy density deposited beyond the beam rapidity, we set $\eta_0 = \mathrm{min}(\eta_0, y_\mathrm{beam} - (y_\mathrm{CM} - y_L))$.
The normalization factor $\mathcal{N}_e (x, y)$ is not a free parameter in our model. It is determined by the local invariant mass $M(x, y)$,
\begin{eqnarray}
   &&\mathcal{N}_e (x, y) = \frac{M(x, y)}{2 \sinh(\eta_0) + \sqrt{\frac{\pi}{2}} \sigma_\eta e^{\sigma_\eta^2/2} C_\eta}  \\
   &&C_\eta = e^{\eta_0}\mathrm{erfc}\left(-\sqrt{\frac{1}{2}} \sigma_\eta\right)  + e^{-\eta_0}\mathrm{erfc}\left(\sqrt{\frac{1}{2}} \sigma_\eta\right).
\end{eqnarray}
Here $\mathrm{erfc}(x)$ is the complementary error function.

\begin{figure}[ht!]
    \centering
    \includegraphics[width=1.0\linewidth]{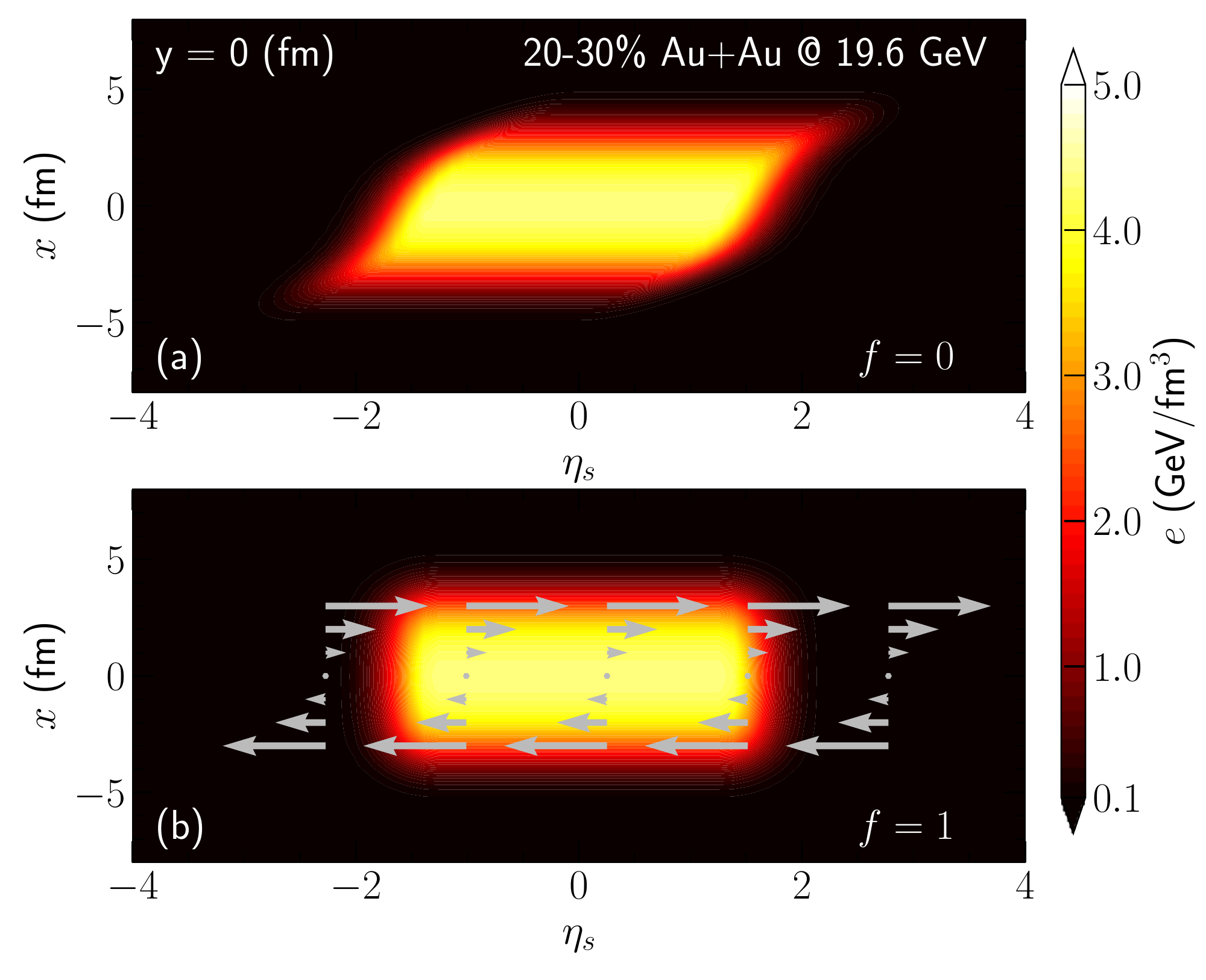}
    \caption{Color contours show the initial energy density distributions in the $x-\eta_s$ plane for 20-30\% Au+Au collisions at 19.6 GeV with the longitudinal rapidity fraction $f = 0$ (a) and $f = 1$ (b). The grey arrows in panel (b) indicate the non-zero initial longitudinal flow $u^\eta$ with $y_L = y_\mathrm{CM}$ in Eqs.~(\ref{eq:Ttautau}) and (\ref{eq:Ttaueta}). $u^\eta = 0$ in panel (a).}
    \label{fig:eprof}
\end{figure}

Figure~\ref{fig:eprof} shows the two extreme scenarios for the energy density and flow distributions of our 3D initial condition for 20-30\% Au+Au collisions at 19.6 GeV with the longitudinal rapidity fraction parameter $f = 0$ and $f = 1$. When $f = 0$, the local net longitudinal momentum leads to a shift of the energy density flux tube to the forward rapidity. While with $f = 1$, the longitudinal momentum $P_z(x, y)$ is attributed to the longitudinal flow velocity instead. Let us note here that ensuring the net longitudinal momentum conservation introduces an anti-correlation between the shifts of the energy density flux tubes in the $\eta_s$ direction and the size of the longitudinal flow velocity. As we will see in the following section, varying the parameter $f$ results in strong dependencies in the $\Lambda$'s polarization and the slope of pion's directed flow $dv_1/dy$.
Therefore, these two experimental observables can tight constraints on the parameter $f$.

In addition to the initial energy and momentum distributions, the non-zero net baryon number current is considered for heavy-ion collisions in the RHIC BES program.
The net-baryon number density current has the form of
\begin{eqnarray}
    J_B^{\mu} (x,y,\eta_s) & = & n_B(x,y,\eta_s) \, u^{\mu}(x,y,\eta_s).
\end{eqnarray}
Here $n_B(x,y,\eta_s)$ represents the local net baryon density
\begin{eqnarray}
    n_B (x,y,\eta_s) & = & T_A (x,y) \, f_{n_B}^{A} (\eta_s) + T_B (x, y) \, f_{n_B}^{B} (\eta_s),\quad
    \label{eq:nBprof}
\end{eqnarray}
where its space-time rapidity dependence is characterized by asymmetric Gaussian functions $f_{n_B}^A$ and $f_{n_B}^B$ as in \cite{Denicol:2018wdp},
\begin{eqnarray}
    f_{n_B}^{A} (\eta_s) & = & \mathcal{N}_{n_B} \left\{ \theta (\eta_s - \eta_{B,0}) \exp{\left[ -\frac{ (\eta_s - \eta_{B,0})^2 }{ 2 \, \sigma_{\scriptsize B, \textrm{out}}^2 } \right]} \right. \nonumber\\
    & & + \left. \theta (\eta_{B,0} - \eta_s) \exp{\left[ -\frac{ (\eta_s - \eta_{B,0})^2 }{ 2 \, \sigma_{\scriptsize B, \textrm{in}}^2 } \right]} \right\} \\
    f_{n_B}^{B} (\eta_s) & = & \mathcal{N}_{n_B} \left\{ \theta (\eta_s + \eta_{B,0}) \exp{\left[ -\frac{ (\eta_s + \eta_{B,0})^2 }{ 2 \, \sigma_{\scriptsize B, \textrm{in}}^2 } \right]} \right. \nonumber\\
    & & + \left. \theta (-\eta_{B,0} - \eta_s) \exp{\left[ -\frac{ (\eta_s + \eta_{B,0})^2 }{ 2 \, \sigma_{\scriptsize B, \textrm{out}}^2 } \right]} \right\}. \quad
\end{eqnarray}
The relevant parameters $\eta_{B,0}$, $\sigma_{\scriptsize B, \textrm{in}}$ and $\sigma_{\scriptsize B, \textrm{out}}$ are determined, such that the net proton rapidity distribution is reproduced \cite{Shen:2020jwv}. We will use the same initial-state model parameters as those in the Table I of Ref.~\cite{Shen:2020jwv} and only vary the new longitudinal momentum fraction parameter $f$ in this work. We have checked that the parameter $f$ has negligible effects on most of the global observables such as the pseudo-rapidity distributions of particle yields, identified particle's mean $p_T$, and elliptic flow coefficient at midrapidity.

\subsection{Hydrodynamic evolution and fluid vorticity}

In this work, we use the open-source 3D viscous hydrodynamic code package \texttt{MUSIC} \cite{Schenke:2010nt, Schenke:2011bn, Paquet:2015lta, Denicol:2018wdp, MUSIC} to simulate fluid dynamical evolution of the system’s energy, momentum, and net baryon density,
\begin{eqnarray}
    \partial_\mu T^{\mu\nu} = 0, \\
    \partial_\mu J_B^\mu = 0,
\end{eqnarray}
where the energy-momentum tensor is defined as
\begin{equation}
    T^{\mu\nu} = e u^\mu u^\nu - (P + \Pi) \Delta^{\mu\nu} + \pi^{\mu\nu}.
\end{equation}
The system's energy-momentum tensor is composed by the local energy density of the fluid cell $e$, the thermal pressure $P$, the fluid velocity $u^\mu$, and the shear stress tensor and bulk viscous pressure $\pi^{\mu\nu}$ and $\Pi$. The spatial projection tensor is defined as $\Delta^{\mu\nu} \equiv g^{\mu\nu} - u^\mu u^\nu$ with the metric $g^{\mu\nu} = diag(1, -1, -1, -1)$. Hydrodynamic equations are solved together with a lattice QCD based Equation of State (EoS) at finite baryon density, \texttt{NEOS-BQS}, in which the strangeness neutrality condition and electric charge density $n_Q = 0.4 n_B$ as imposed \cite{Monnai:2019hkn}. 

In this work, we do not consider viscous effects from bulk viscous pressure, $\Pi = 0$, nor the net baryon diffusion effects. The shear stress tensor is evolved according to the following equation of motion \cite{Denicol:2012cn},
\begin{eqnarray}
    \tau_{\pi} D \pi^{\langle \mu\nu \rangle}
	+ \pi^{\mu\nu}
	& = & 2\eta\,\sigma^{\mu\nu}
	- \delta_{\pi \pi} \pi^{\mu\nu} \theta
	+ \varphi_7 \pi_{\alpha}^{\langle \mu} \pi^{\nu\rangle \alpha} \nonumber \\
	& & - \tau_{\pi \pi} \pi_{\alpha}^{\langle \mu} \sigma^{\nu\rangle \alpha}
	+ \lambda_{\pi \Pi} \Pi\, \sigma^{\mu\nu}\,.
	\label{eq:shear}    
\end{eqnarray}
Here $D = u^\alpha \partial_\alpha$ is the comoving time derivative and $A^{\langle \mu \nu \rangle} = \Delta^{\mu\nu}_{\alpha \beta} A^{\alpha \beta}$ denotes symmetrized and traceless projections with
\begin{eqnarray}
    \Delta^{\mu\nu}_{\alpha \beta} = \frac{1}{2}(\Delta^\mu\,_\alpha \Delta^\nu\,_\beta + \Delta^\nu\,_\alpha \Delta^\mu\,_\beta) - \frac{1}{3} \Delta^{\mu\nu} \Delta_{\alpha \beta}.
\end{eqnarray}
In Eq.~(\ref{eq:shear}), $\eta$ denotes the shear viscosity and $\tau_\pi$ is the relaxation time, which controls the time scale for the shear stress tensor to relax to its Navier-Stokes value. The velocity shear tensor is defined as $\sigma^{\mu\nu} \equiv \frac{1}{2}(\nabla^\mu u^\nu + \nabla^\nu u^\mu) - \frac{1}{3} \Delta^{\mu\nu} (\nabla \cdot u)$, where $\nabla^\mu = \Delta^{\mu \alpha} \partial_\alpha$. Additional second-order gradient terms are included with their transport coefficients $\{\delta_{\pi\pi}, \phi_7, \tau_{\pi\pi}, \lambda_{\pi\Pi}\}$ according to the DNMR hydrodynamic theory \cite{Denicol:2012cn, Denicol:2014vaa}. We use a temperature and $\mu_B$ dependent specific shear viscosity $(\eta/s)(T, \mu_B)$ in our hydrodynamic simulations as in Ref.~\cite{Shen:2020jwv}. This $(\eta/s)(T, \mu_B)$ is constrained by the elliptic flow measurements from the RHIC BES phase I \cite{Adamczyk:2017hdl}.

During hydrodynamic simulations, the fluid kinematic vorticity tensor can be computed as,
\begin{equation}
    \omega^{\mu\nu}_{K} \equiv \frac{1}{2} \left(\partial^\nu u^\mu - \partial^\mu u^\nu \right).
    \label{eq:omega_K}
\end{equation}
One can also define the transverse kinematic vorticity tensor with the spatial projection operator,
\begin{equation}
    \omega^{\mu\nu}_{K,\perp} \equiv \frac{1}{2}\left(\nabla^\nu u^\mu - \nabla^\mu u^\nu \right),
    \label{eq:omega_KSP}
\end{equation}
The transverse kinematic vorticity differs from the kinematic vorticity tensor by the local acceleration,
\begin{eqnarray}
    \omega^{\mu\nu}_{K,\perp} &\equiv& \frac{1}{2}(\partial^\nu u^\mu - \partial^\mu u^\nu) - \frac{1}{2}(u^\nu D u^\mu - u^\mu D u^\nu)
    \nonumber \\
    &=& \omega^{\mu\nu}_{K} - \frac{1}{2}(u^\nu D u^\mu - u^\mu D u^\nu).
\end{eqnarray}
The thermal vorticity is defined as
\begin{eqnarray}
    \omega^{\mu\nu}_\mathrm{th} &\equiv& \frac{1}{2} \left[\partial^\nu \left(\frac{u^\mu}{T}\right) - \partial^\mu \left( \frac{u^\nu}{T} \right) \right]
    \label{eq:omega_thermal} \label{eq:omega_thermal} \\
    &=& \frac{1}{T} \left\{\omega^{\mu\nu}_{K} - \frac{1}{2T} [(\partial^\nu T) u^\mu - (\partial^\mu T) u^\nu] \right\} \nonumber
\end{eqnarray}
and the $T$-vorticity is
\begin{eqnarray}
    \omega^{\mu\nu}_T &\equiv& \frac{1}{2} \left(\partial^\nu (T u^\mu) - \partial^\mu (T u^\nu) \right)
    \label{eq:omega_T} \\
    &=& T \left\{ \omega^{\mu\nu}_{K} + \frac{1}{2 T} [(\partial^\nu T) u^\mu - (\partial^\mu T) u^\nu] \right\} \nonumber
\end{eqnarray}
The thermal and $T$-vorticity tensors receive opposite contribution from the temperature gradient terms. We will explore the theoretical uncertainty of computing the hyperon's spin polarization with different types of vorticity tensors in Appendix~\ref{sec:appendixA}.

\subsection{Evolution of the fluid vorticity near midrapidity}

We define the collision impact parameter along the $+x$ direction and points from the target nucleus to the projectile. In this convention, the global OAM points to the $-y$ direction. The $\Lambda$ hyperon's global polarization is defined as its polarization component along the global OAM direction, which is related to the $xz$ component of the thermal vorticity tensor $\omega^{\mu\nu}_\mathrm{th}$. It is instructive first to study the time evolution of $\omega_\mathrm{th}^{xz}$ during the hydrodynamic evolution. We define the thermal vorticity averaged over a given space-time volume weighted by the local energy density,
\begin{equation}
    \langle \omega_\mathrm{th}^{\mu\nu} \rangle (\tau) = \frac{\int^{\eta_{s}^\mathrm{max}}_{\eta_{s}^\mathrm{min}} d\eta_s \int d^2 x_\perp e \omega_\mathrm{th}^{\mu\nu} }{\int^{\eta_{s}^\mathrm{max}}_{\eta_{s}^\mathrm{min}} d \eta_s \int d^2 x_\perp e}.
\end{equation}
For midrapidity fluid cells, we choose a symmetric space-time rapidity window, $\eta_s^\mathrm{min} = -0.5$ and $\eta_s^\mathrm{max} = 0.5$.

\begin{figure}[t!]
    \centering
    \includegraphics[width=0.95\linewidth]{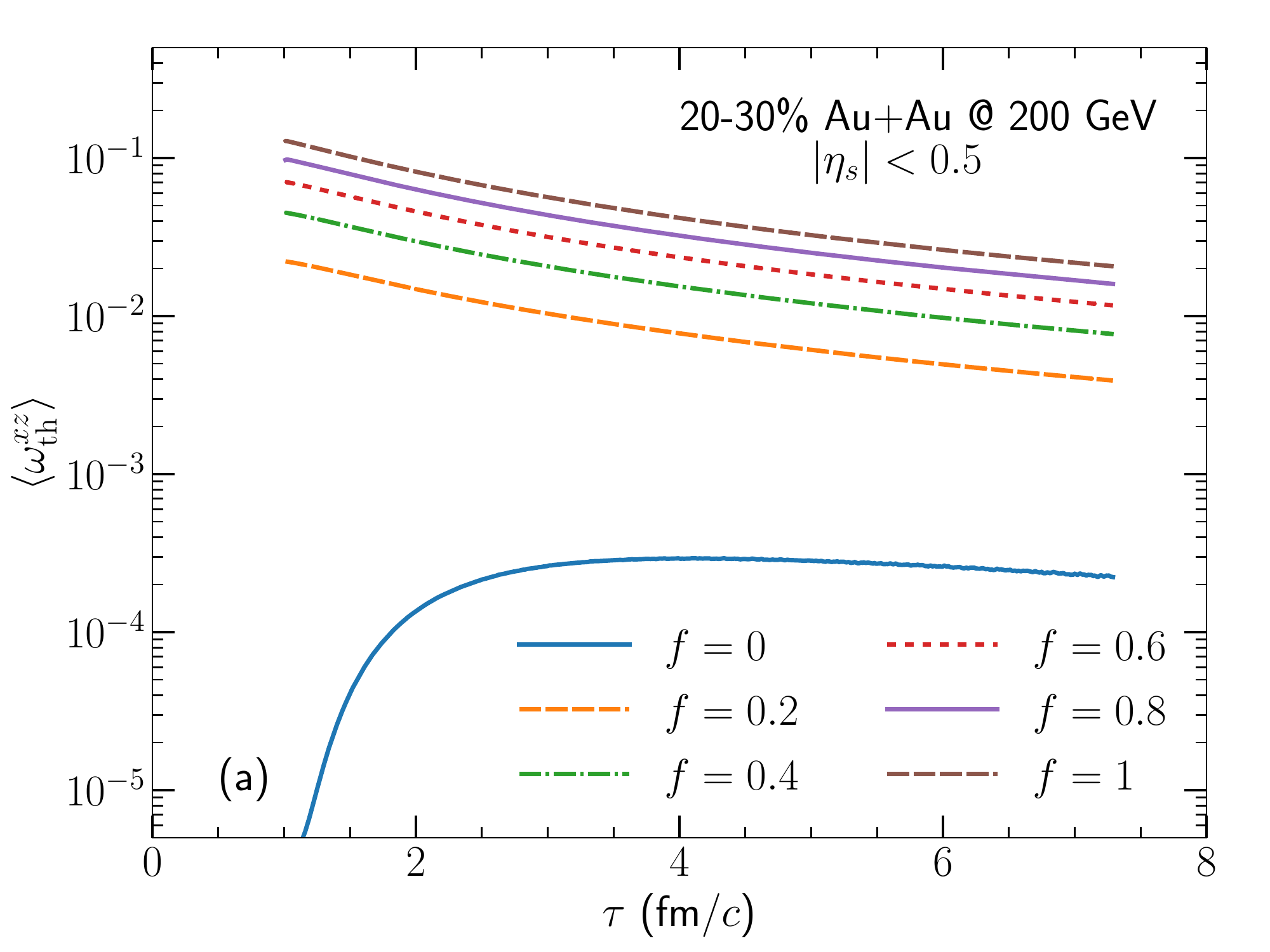}
    \includegraphics[width=0.95\linewidth]{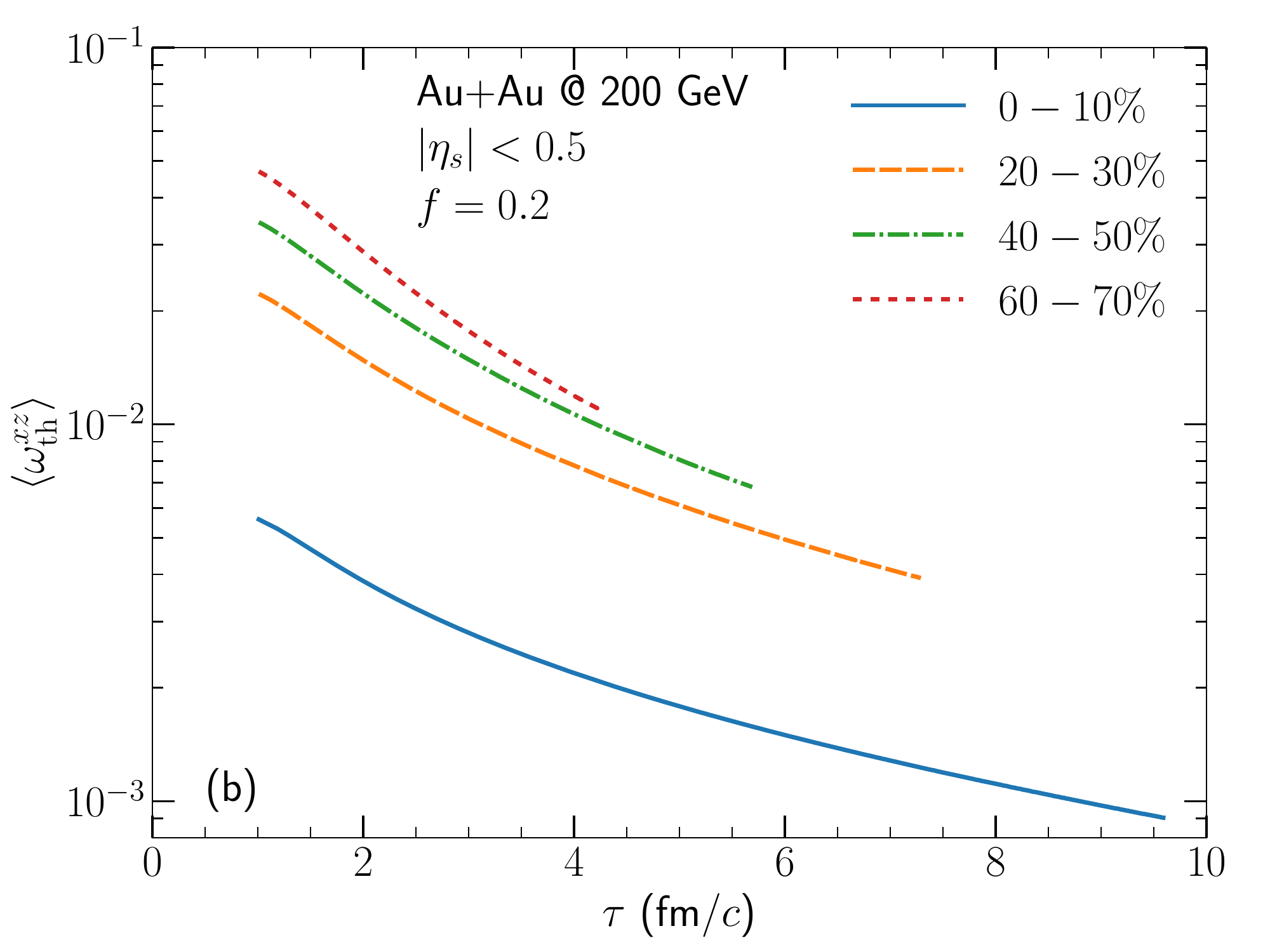}
    \caption{(Color online) Panel (a): Time evolution of the averaged thermal vorticity of fluid with different longitudinal rapidity fraction $f$ in mid-rapidity 20-30\% Au+Au collisions at 200 GeV. Panel (b): Time evolution of the averaged thermal vorticity of fluid for four centrality bins in Au+Au collisions at 200 GeV with $f = 0.2$.}
    \label{fig:vor_evo}
\end{figure}

As Fig.~\ref{fig:eprof} illustrated, the longitudinal rapidity fraction parameter $f$ controls how much of the global OAM is attributed to the initial local fluid vorticity. We find that the initial averaged fluid vorticity $\langle \omega_\mathrm{th}^{\mu\nu} \rangle$ has a good linear dependence on the model parameter $f$.

Figure~\ref{fig:vor_evo}a shows the evolution of the averaged fluid vorticity in 20-30\% Au+Au collisions at 200 GeV with different values  of $f$. With the parameter $f = 0$, all the system's OAM is attributed to the shifts of energy density flux tubes along the $\eta_s$ direction. The entire system starts with zero fluid vorticity $ \omega_\mathrm{th}^{xz}$ at the beginning of hydrodynamic simulations. We observe that the averaged $\langle \omega_\mathrm{th}^{xz} \rangle$ increases rapidly during the first fm/$c$ of the hydrodynamic evolution and saturates around with a magnitude of $10^{-4}$ afterward. Our result suggests that the pressure gradients inside the fluid can develop vorticity within a time-scale of 1 fm/$c$, but the size is small at 200 GeV. With a non-zero $f$ value in the initial-state model, a fraction of the OAM is attributed to non-zero initial fluid vorticity. In these cases, the averaged $\langle \omega_\mathrm{th}^{xz} \rangle$ decreases monotonically as a function of $\tau$. This qualitatively different time evolution between $f = 0$ and $f \neq 0$ indicates that the initial-state longitudinal flow distribution dominates the fluid thermal vorticity $\omega^{xz}_\mathrm{th}$ (related to the global polarization) in heavy-ion collisions.

Figure~\ref{fig:vor_evo}b shows the evolution of the averaged fluid vorticity $\langle \omega_\mathrm{th}^{xz} \rangle(\tau)$ in four centrality bins in Au+Au collisions at 200 GeV. With all the model parameter fixed, the initial fluid vorticity is larger in the more peripheral centrality bin. This centrality dependence is because of the large local asymmetry between $T_A$ and $T_B$ in the peripheral collisions. The time evolution of $\langle \omega_\mathrm{th}^{xz} \rangle(\tau)$ is qualitatively the same for all centrality bins in the hydrodynamic phase. Our results have qualitatively the same behavior as those in the transport models \cite{Jiang:2016woz}. 

\subsection{The averaged spin vector of fermions}

For spin-1/2 fermions, the average spin vector (defined as the Pauli-Lubanski vector) over the hyper-surface $\Sigma_\mu$ can be computed as \cite{Liu:2021uhn}
\begin{eqnarray}
    S^\mu(p^\mu) = \frac{1}{4m} \frac{\int d^3 \Sigma_\alpha p^\alpha \mathcal{A}^\mu}{\int d^3 \Sigma_\alpha p^\alpha n_0(E)}.
    \label{eq:SpinVector}
\end{eqnarray}
Here, the axial vector is defined as,
\begin{eqnarray}
    \mathcal{A}^\mu &=& \beta n_0(E)(1 - n_0(E)) \nonumber \\
    && \times \epsilon^{\mu\nu\alpha\gamma} \bigg[ -\frac{1}{2\beta} p_\nu \omega_{\alpha \gamma}^\mathrm{th} - \frac{b_i}{\beta E} u_\nu p_{\perp \alpha} \nabla_\gamma \frac{\mu_B}{T} \nonumber \\
    && \qquad \qquad - \frac{p_\perp^2}{E} u_\nu Q_\alpha\,^\rho \sigma_{\rho\gamma} \bigg],
    \label{eq:axialVector}
\end{eqnarray}
where $E = p^\mu u_\mu$, $p^\mu_\perp = \Delta^{\mu\nu} p_\nu$, and $Q^{\mu\nu} = - \frac{p_\perp^\mu p_\perp^\nu}{p_\perp^2} + \frac{1}{3} \Delta^{\mu\nu}$.
Here $\epsilon^{\mu \rho \sigma \tau}$ is the Levi-Civita tensor and we choose the convention $\epsilon^{txyz} = 1$. We denote the term related to $\nabla_\gamma (\mu_B/T)$ as the $\mu_B$ Induced Polarization ($\mu_B$IP) \cite{Liu:2020dxg} and the last term related to the velocity shear tensor as the Shear Induced Polarization (SIP)\footnote{We notice that the shear-induced Polarization term has a different expression in \cite{Becattini:2021suc}, where $u_\nu$ was replaced by a global time vector $t_\nu = (1, 0, 0, 0)$ and the $\sigma_{\rho\gamma}$ included additional temperature gradients. While the exact form of the SIP is still under debate, we will carry out calculations with the SIP definition in Eq.~(\ref{eq:axialVector}) in this work. Our conclusions do not depend on the exact forms of the SIP term.} \cite{Becattini:2021suc, Liu:2021uhn}.
Equations~(\ref{eq:SpinVector}) and (\ref{eq:axialVector}) assume that the hyper-surface fluid cells reach local thermal equilibrium. The fermions emitted at early-time of the evolution could receive sizable  out-of-equilibrium corrections.

In this work, we compute $\Lambda$ and $\bar{\Lambda}$'s spins on a constant energy hyper-surface with $e = e_\mathrm{sw}$, on which fluid cells are converted to hadrons via the Cooper-Frye prescription. Hadrons are further fed to the \texttt{UrQMD} hadronic transport. Because \texttt{UrQMD} does not distinguish hadrons' spins in their evolution, we assume the spins of $\Lambda$ and $\bar{\Lambda}$ are frozen-out at $e = e_\mathrm{sw}$ in this work. The values of $e_\mathrm{sw}$ are adjusted to match the proton yield in every collision energy at the RHIC BES program \cite{Oliinychenko:2020znl}. We will study how our results depend on the choice of $e_\mathrm{sw}$ in Appendix~\ref{sec:appendixB}.

The averaged polarization vector in the lab frame is
\begin{eqnarray}
    P_\mathrm{lab}^\mu(p^\mu) = S^\mu(p^\mu)/\langle S \rangle. \label{eq:polarizationVec}
\end{eqnarray}
In the RHIC experiments, the polarizations of $\Lambda$ and $\bar{\Lambda}$ are measured in the particle's local rest frame,
\begin{equation}
    P^t(p^\mu) = \frac{p^0}{m} P_\mathrm{lab}^t(p^\mu) - \frac{\vec{p} \cdot \vec{P}_\mathrm{lab}(p^\mu)}{m} = 0
\end{equation}
and
\begin{equation}
    P^i(p^\mu) = P_\mathrm{lab}^i(p^\mu) - \frac{\vec{p} \cdot \vec{P}_\mathrm{lab}(p^\mu)}{p^0 (p^0 + m)} p^i.
    \label{eq:P_LRF}
\end{equation}
In the $\Lambda$'s local rest frame, the time component of $P^\mu$ is zero, which serves as a non-trivial test for the numerical implementations.

It is instructive to understand the time development of $\Lambda$ hyperon's polarization during hydrodynamic evolution. Based on Eqs.~(\ref{eq:SpinVector}) and (\ref{eq:polarizationVec}), we can compute the differential polarization vector as a function of the hydrodynamic proper time $\tau$,
\begin{eqnarray}
    P_\mathrm{lab}^\mu(p^\mu, \tau) = \lim_{\Delta \tau \rightarrow 0} \frac{1}{\langle S \rangle} \frac{1}{4m} \frac{ \int_{\tau}^{\tau + \Delta \tau} d^3 \Sigma_\alpha p^\alpha \mathcal{A}^\mu}{\int_{\tau}^{\tau + \Delta \tau} d^3 \Sigma_\alpha p^\alpha n_0(E)}. \nonumber \\ 
    \label{eq:dPdtau}
\end{eqnarray}
We then boost the $P_\mathrm{lab}^\mu(p^\mu, \tau)$ to the hyperon's local rest frame with Eq.~(\ref{eq:P_LRF}) and denote it as $P^\mu(p^\mu, \tau)$. Please note that we normalize the differential polarization vector by the number of hyperon emitted within the $\Delta \tau$ interval,
\begin{equation}
    \frac{dN}{d\tau}(p^\mu, \tau) = \lim_{\Delta \tau \rightarrow 0} \frac{1}{\Delta \tau} \int_{\tau}^{\tau + \Delta \tau} d^3 \Sigma_\alpha p^\alpha n_0(E).
\end{equation}
The momentum-integrated hyperon polarization at time $\tau$ can be computed as a yield-weighted average,
\begin{equation}
    P^\mu(\tau) = \frac{\int \frac{d^3p}{E} P^\mu(p^\mu, \tau) \frac{dN}{d\tau}(p^\mu, \tau)}{\int \frac{d^3p}{E} \frac{dN}{d\tau}(p^\mu, \tau)}.
\end{equation}
To study $P^\mu(\tau)$'s contribution to the total hyperon polarization, we need to weight $P^\mu(\tau)$ with the number of hyperon emitted at every time step $\tau$,
\begin{equation}
    \frac{\Delta P^\mu}{\Delta \tau}(\tau) = \frac{P^\mu(\tau) \int \frac{d^3p}{E} \frac{dN}{d\tau}(p^\mu, \tau)}{\int d\tau \int \frac{d^3p}{E} \frac{dN}{d\tau}(p^\mu, \tau)}.
    \label{eq:dPdtau}
\end{equation}

\begin{figure}[ht!]
    \centering
    \includegraphics[width=0.95\linewidth]{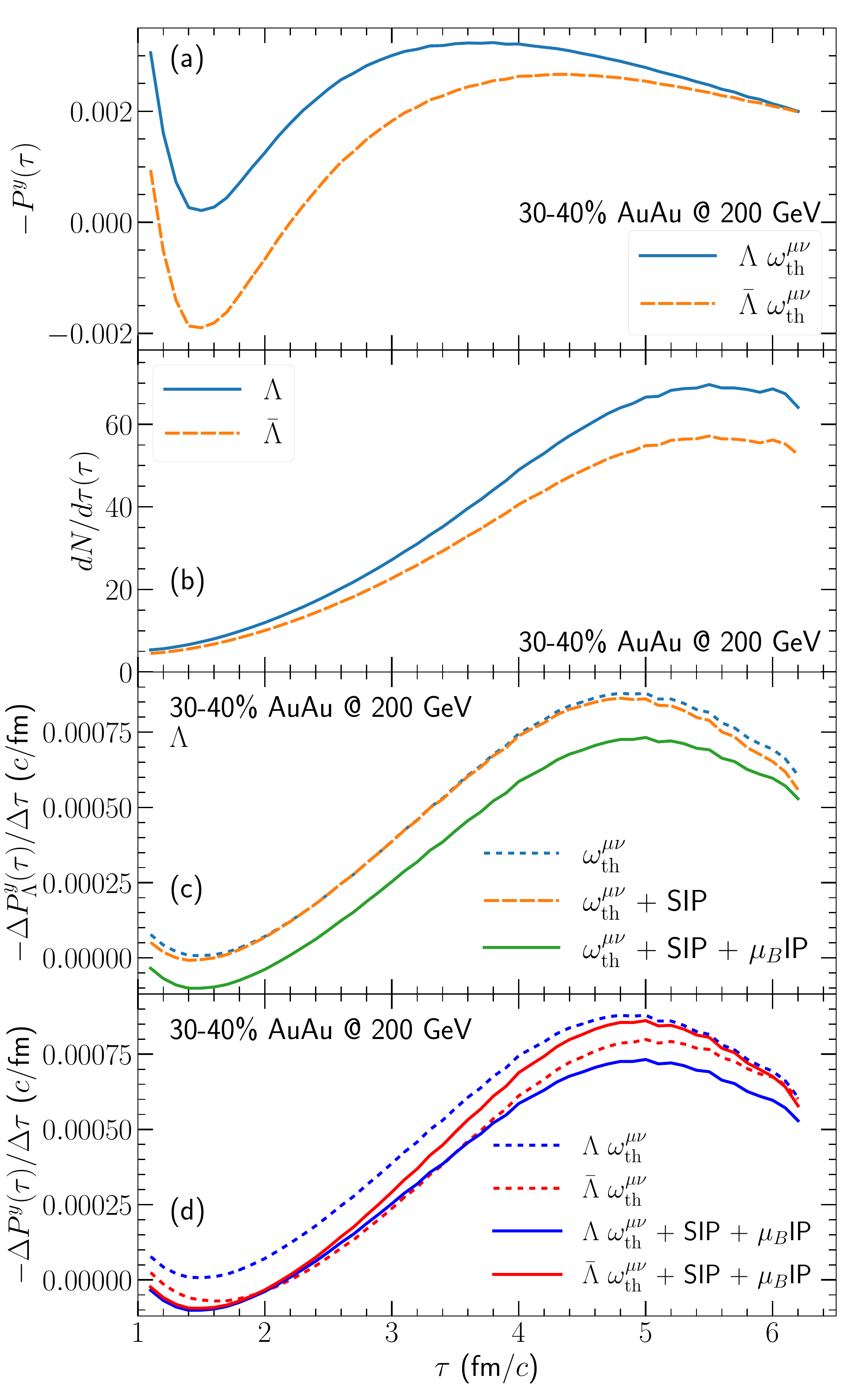}
    \caption{(Color online) Panel (a): The hyperon's global polarization as a function of hydrodynamic proper time. Panel (b): The hyperon production as a function of $\tau$. Panel (c): The time development of $\Lambda$'s global polarization with different fluid gradients. Panel (d): The comparison of $\Lambda$ and $\bar{\Lambda}$'s global polarization developments. The results are for $\Lambda$ and $\bar{\Lambda}$ with $p_T \in [0.5, 3.0]$ GeV and $\vert y \vert < 1$ in 30-40\% Au+Au collisions at 200 GeV with the longitudinal rapidity fraction $f = 0.2$.}
    \label{fig:Pyevo}
\end{figure}

Figure~\ref{fig:Pyevo}a shows that the averaged hyperon polarization as a function of the longitudinal proper time. The $P^y(\tau)$ drops sharply during the first 0.5 fm/$c$, following the evolution of averaged $\langle \omega^{xz}_\mathrm{th} \rangle$ in Fig.~\ref{fig:vor_evo}. Then $P^y(\tau)$ gradually increases and reaches its peak around 2.5 fm/$c$ in the hydrodynamic evolution, which is from the $\omega^{tx}_\mathrm{th}$'s contribution in Eq.~(\ref{eq:axialVector}). Figure~\ref{fig:Pyevo}b shows the hyperon production is dominated by the time-like surface elements (enhanced by the $\tau$ factor in the Jacobian) in the Cooper-Frye particlization at late time.  By weighting $P^y(\tau)$ with the number of hyperons emitted at every time step in Eq.~(\ref{eq:dPdtau}), we find that most contributions to the total polarization come from late time of the hydrodynamic evolution, as shown in Figs.~\ref{fig:Pyevo}c and \ref{fig:Pyevo}d. Although the early-time emitted hyperons are also largely polarized and could receive sizable out-of-equilibrium corrections, their net contributions to the total polarization remain small. 
Figure~\ref{fig:Pyevo}c demonstrates the effects of different fluid gradients in Eq.~(\ref{eq:axialVector}) on the development of $\Lambda$'s global polarization during hydrodynamic evolution. The thermal vorticity gives the dominant contribution to $\Lambda$'s global polarization.
The contribution of shear-induced polarization (SIP) to the integrated global polarization is negligible as expected from its tensor structure in Eq.~(\ref{eq:axialVector}). The $\mu_B/T$ gradients suppress the $\Lambda$'s global polarization by roughly a constant over time.

Figure~\ref{fig:Pyevo}d further compares the time development of $\Lambda$ and $\bar{\Lambda}$'s global polarization in 30-40\% Au+Au collisions at 200 GeV. With the non-zero baryon density in the fluid, $\Lambda$ hyperons receive larger contributions to their global polarization from the fluid thermal vorticity than those to $\bar{\Lambda}$. This effect is caused by the $\mu_B$'s dipolar transverse distribution in the forward and backward space-time rapidities, which imprint the shapes of the projectile and target nuclei's nuclear thickness functions as in Eq.~(\ref{eq:nBprof}).
The $\mu_B$ gradient-induced polarization ($\mu_B$IP) gives opposite contributions to $\Lambda$ and $\bar{\Lambda}$. It cancels the difference between $\Lambda$ and $\bar{\Lambda}$ during the first two fm/$c$ of the evolution and contributes more to $\bar{\Lambda}$ in the late stage.

\section{Polarization Results at the RHIC BES program}
\label{sec:polarization_results}

Before we compare our calculations of the $\Lambda$ and $\bar{\Lambda}$'s global polarization with the RHIC BES measurements, it is essential to understand the effects of the longitudinal rapidity fraction parameter $f$ on various experimental observables. On the one hand, we checked that this model parameter does not have noticeable effects on particle rapidity distribution, mean transverse momentum, nor elliptic flow coefficient at mid-rapidity. On the other hand, it shows strong sensitivity to the $\Lambda$'s global polarization and the slope of rapidity dependent $\pi$'s directed flow, $dv_1/dy\vert_{y=0}$. These two experimental observables are sensitive probes to the initial longitudinal flow and the energy density's space-time rapidity distribution.

\begin{figure}[ht!]
    \centering
    \includegraphics[width=\linewidth]{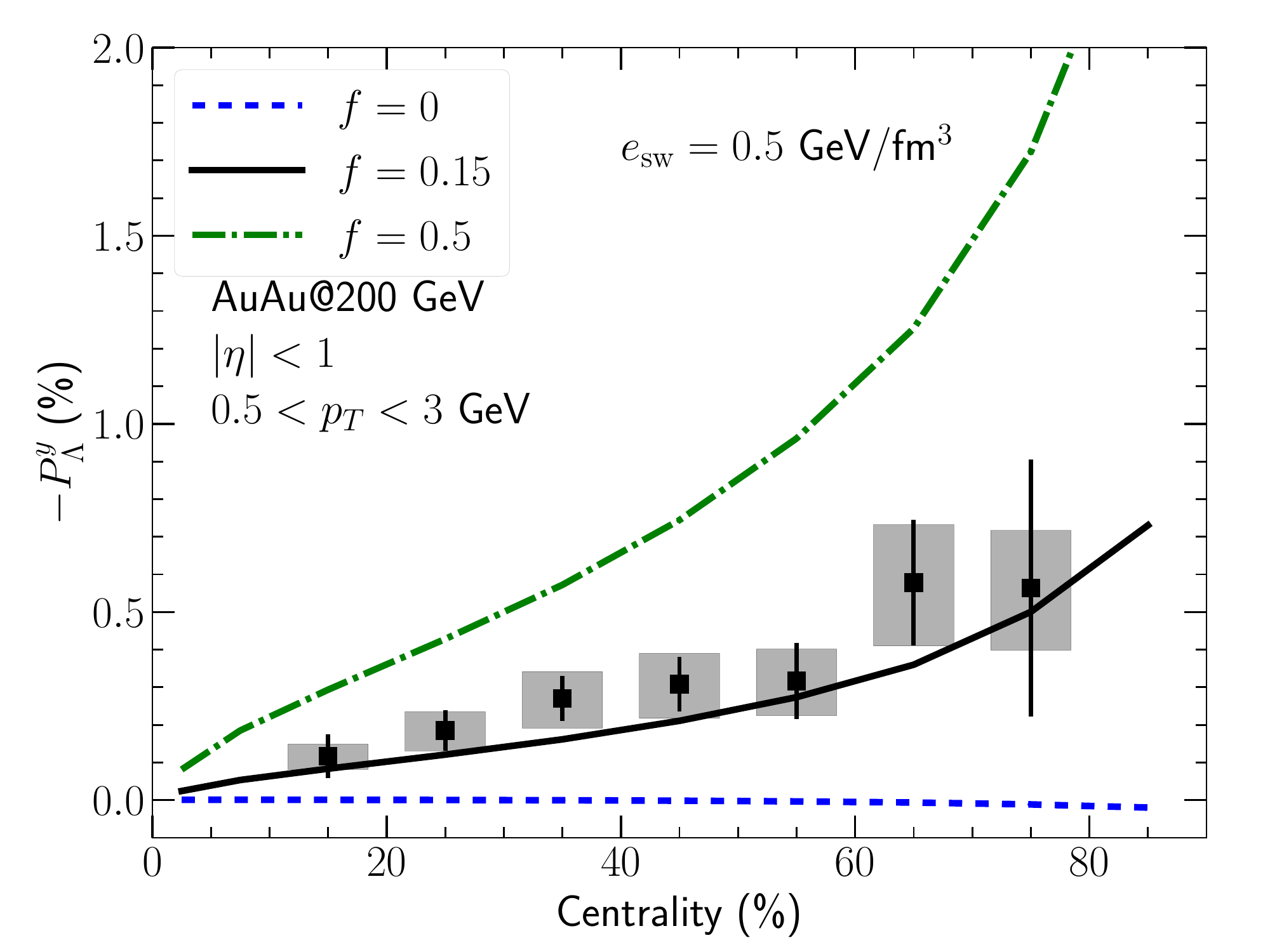}
    \includegraphics[width=0.95\linewidth]{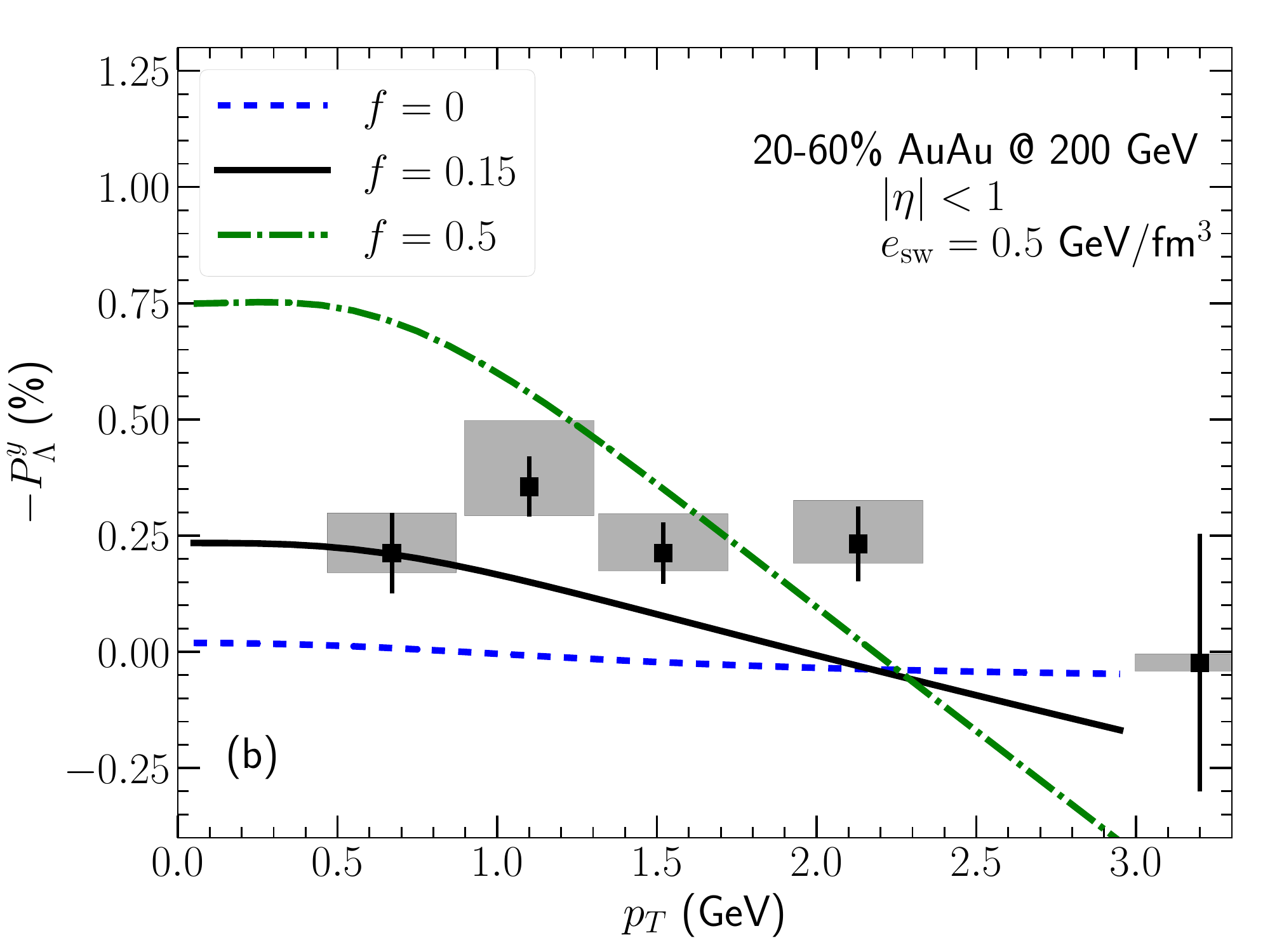}
    \includegraphics[width=\linewidth]{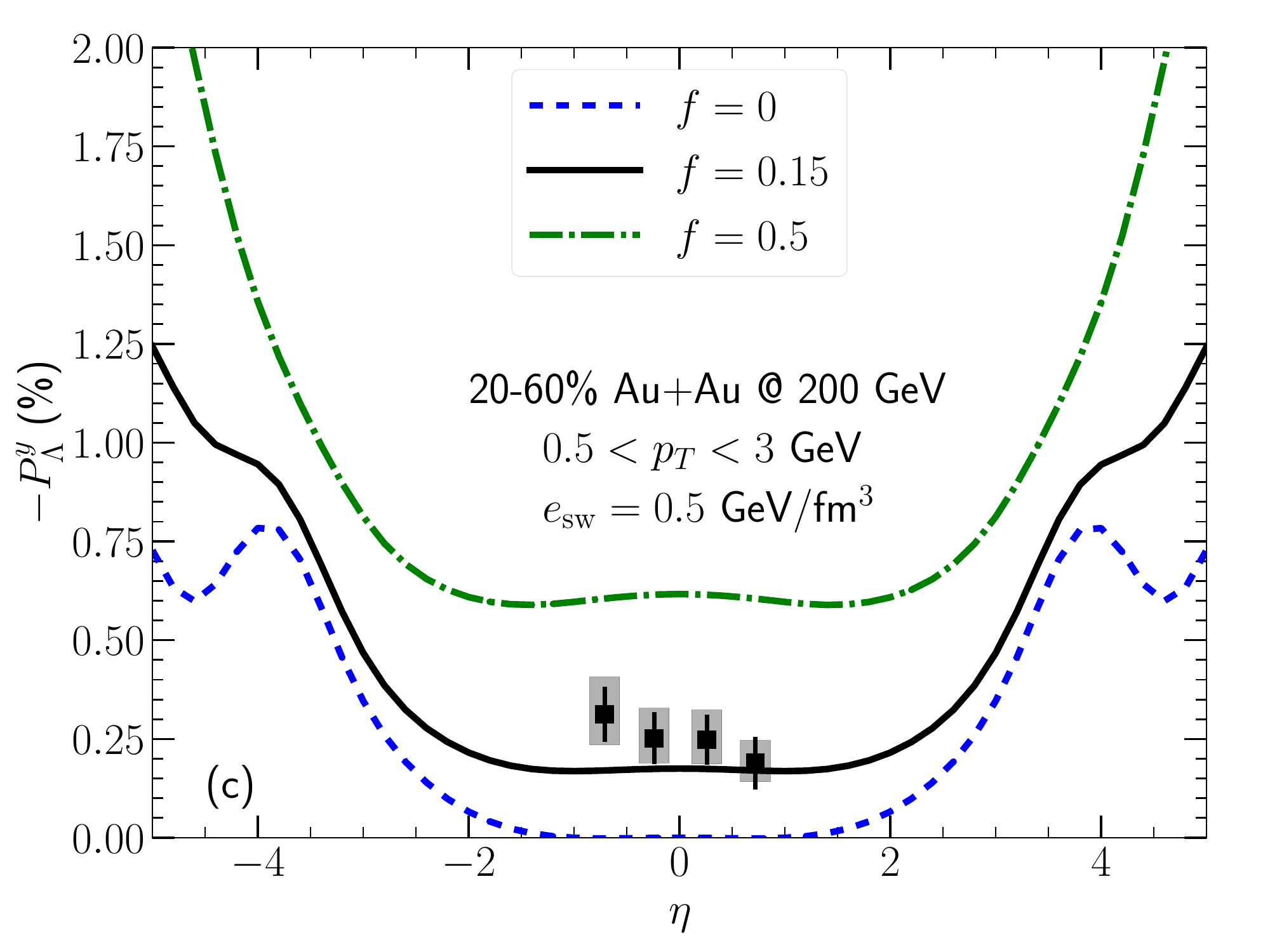}
    \caption{(Color online) The global $\Lambda$ polarization's dependence on the initial-state longitudinal rapidity fraction in Au+Au collisions at 200 GeV compared with the STAR measurements \cite{Adam:2018ivw}. The $\Lambda$'s global polarization is computed with all the gradient terms in Eq.~(\ref{eq:axialVector}). Panel (a) shows the $P^y_\Lambda$'s centrality dependence. Panel (b) presents the $p_T$-differential $P^y_\Lambda$ in 20-60\% Au+Au collisions. Panel (c) shows the pseudo-rapidity dependence of $P^y_\Lambda$.}
    \label{fig:Au200_yLfracDep}
\end{figure}

Figure~\ref{fig:Au200_yLfracDep} shows that the magnitudes of $\Lambda$'s global polarization are very sensitive to the value of the longitudinal rapidity fraction parameter $f$ in our model. With $f = 0$, the entire fluid starts with zero $\omega^{xz}$ at the beginning of hydrodynamics.
The $P^y_\Lambda$ remains almost zero in the mid-rapidity region, which is expected from the thermal vorticity evolution shown in Fig.~\ref{fig:vor_evo}. We find a constant $f = 0.15$ can give a good description of the centrality dependence of the $P^y_\Lambda$ in Au+Au collisions at 200 GeV, while the results with $f = 0.5$ already overestimate the STAR measurements by a factor of two.  Figure~\ref{fig:Au200_yLfracDep}b shows the global polarization decreases monotonically as a function of $p_T$. Due to the presence of the thermal distribution $n_0 (p \cdot u)$ in the expression for the polarization, one can also anticipate that the global spin polarization can receive significant contribution from $\mathcal{A}^{\mu}$ at low momentum.
At zero transvese momentum limit $p^\mu = (m, 0, 0, 0)$,
\begin{eqnarray}
    P^y = P^y_\mathrm{lab} \propto \mathcal{A}^y &=& n_0(m)(1 - n_0(m))\bigg[ - m \omega^{xz}_\mathrm{th} \nonumber \\
    && - b_i \bigg( -u^x \partial^z \frac{\mu_B}{T} + u^z \partial^x \frac{\mu_B}{T} \bigg) \nonumber \\
    && + \frac{m}{T} ( -u^x \sigma^{tz} + u^z \sigma^{tx}) \bigg].
\end{eqnarray}
We have checked that the dominant numerical contribution comes from the thermal vorticity tensor $\omega^{xz}_\mathrm{th}$. Therefore, the global polarization at zero transverse momentum is directly related to the fluid thermal vorticity component $\omega^{xz}_\mathrm{th}$, recovering the non-relativistic limit. While for finite $p_T$, the $\omega_\mathrm{th}^{tx}$ and $\omega_\mathrm{th}^{tz}$ give additional relativistic contributions to $\Lambda$'s polarization.
A larger longitudinal rapidity fraction $f$ in the initial condition results in a larger global polarization $P^y$ at $p_T = 0$ and a steeper decrease as $p_T$ increases. 

Finally, Figure~\ref{fig:Au200_yLfracDep}c shows the pseudo-rapidity dependence of $P^y_\Lambda$. In semi-peripheral Au+Au collisions at 200 GeV, the polarization $P^y_\Lambda$ has a plateau for $\vert \eta \vert < 2$ and increases in the forward and backward rapidity regions. Different values of $f$ shift the magnitude of $P^y(\eta)$ by constants for $\vert \eta \vert < 2$.

\begin{figure}[ht!]
    \centering
    \includegraphics[width=0.95\linewidth]{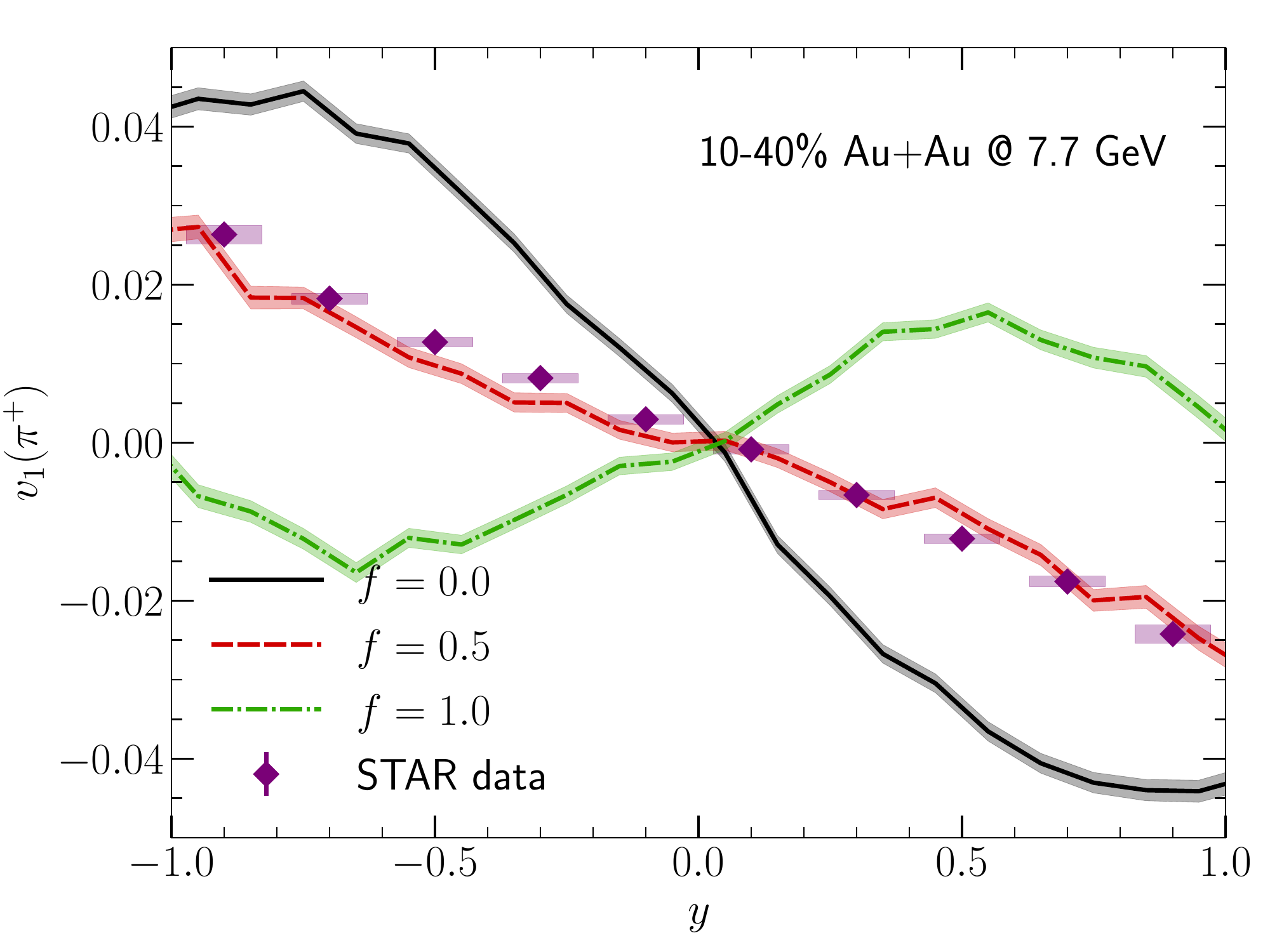}
    \caption{(Color online) The directed flow of $\pi^+$ as a function of rapidity with different initial-state longitudinal rapidity fraction $f$ for 10-40\% Au+Au collisions at 7.7 GeV compared with the STAR measurement \cite{Adamczyk:2014ipa}.}
    \label{fig:v1slope}
\end{figure}

Figure~\ref{fig:v1slope} shows a strong positive correlation between the slope of pion's directed flow and the initial longitudinal rapidity fraction parameter $f$ in our model. As the value of $f$ varies from 0 to 1 in the model, there are fewer longitudinal shifts of initial energy density distribution as shown in Fig.~\ref{fig:eprof}, which result in a reduction of dipolar transverse deformation in the initial energy density profile in forward and backward space-time rapidities. Therefore, simulations with a large $f$ value give a small slope for the pion's directed flow $dv_1/dy$ at mid-rapidity. We find that $f = 0.5$ is preferred for Au+Au collisions at 7.7 GeV compared with the STAR measurements. The positive $dv_1/dy$ in the $f = 1$ case is generated by the dipolar deformation of the initial state net baryon density in the calculation.

\begin{figure}[t!]
    \centering
    \includegraphics[width=0.95\linewidth]{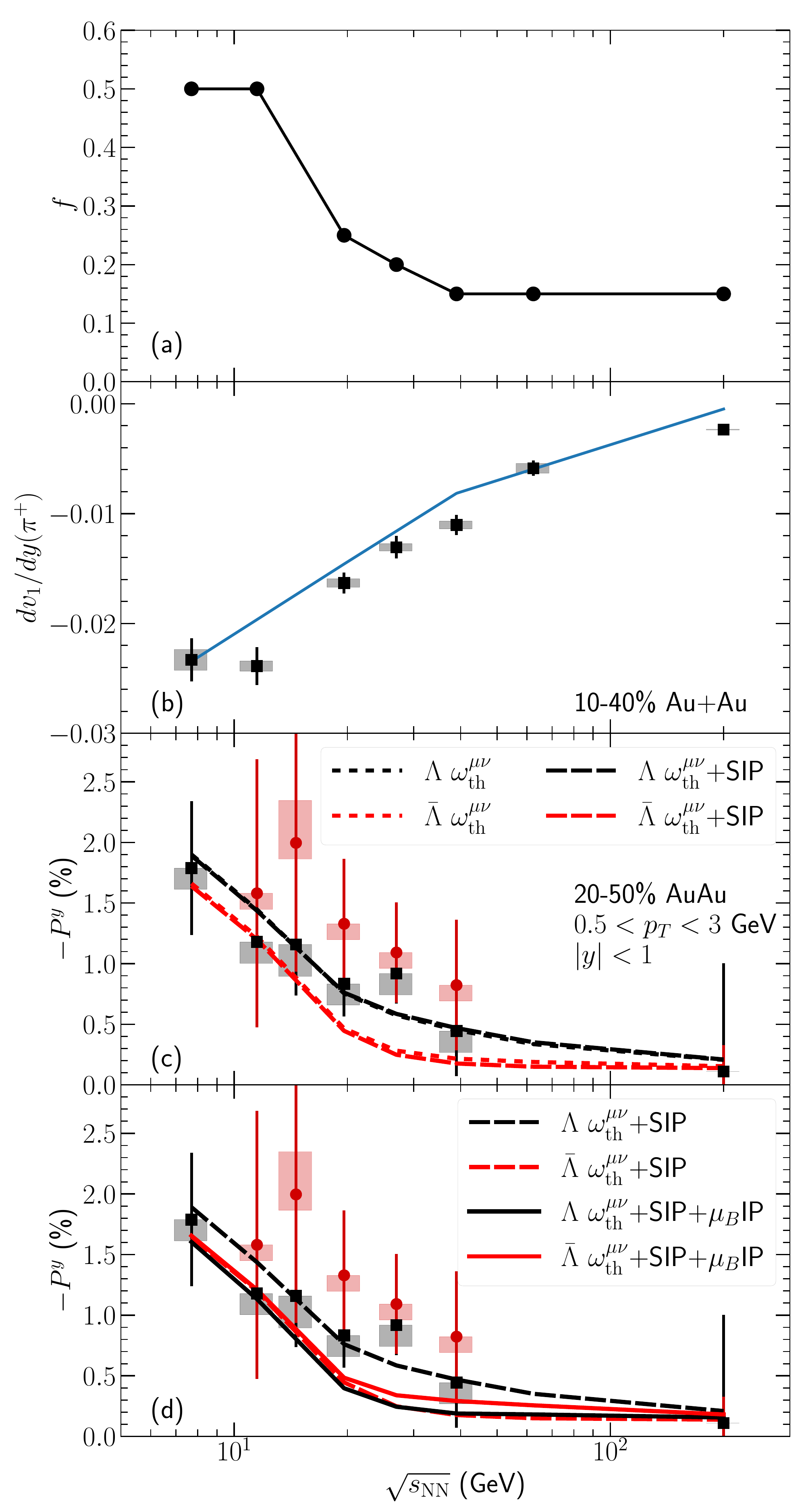}
    \caption{(Color online) Panel (a): The value of longitudinal rapidity fraction $f$ as a function of collision energy. Panel (b): The slope of $\pi^+$ directed flow at $y = 0$, $dv_1/dy \vert_{y = 0}$, compared with the STAR measurements \cite{Adamczyk:2014ipa}. Panels (c) and (d): The global $\Lambda$ polarization in 20-50\% Au+Au collisions as a function of collision energy. Calculations including different gradient terms are compared with the STAR measurements \cite{STAR:2017ckg}. The STAR polarization data points are rescaled by 0.877 because the latest hyperon decay parameter $\alpha_\Lambda$ \cite{Zyla:2020zbs}.}
    \label{fig:Py_vs_ecm}
\end{figure}

\begin{figure}[ht!]
    \centering
    \includegraphics[width=0.95\linewidth]{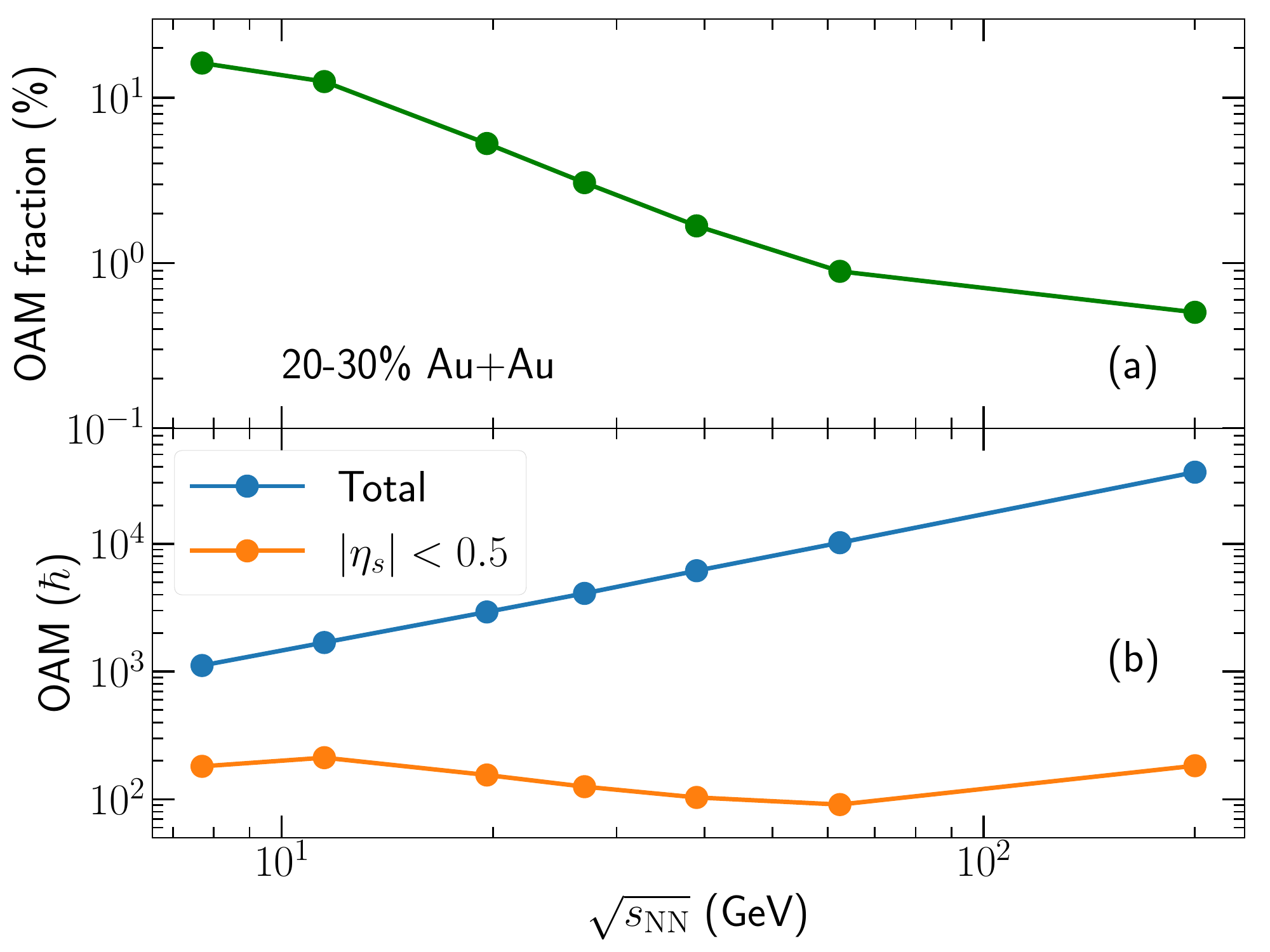}
    \caption{(Color online) Panel (a): The percentage fraction of orbital angular momentum (OAM) in the mid-rapidity region $\vert \eta_s \vert < 0.5$ relative to the system's total OAM from the participant nucleons for 20-30\% Au+Au collisions at the RHIC BES energies. Panel (b): The system's total and mid-rapidity OAM as a function of the collision energy.}
    \label{fig:OAMfraction}
\end{figure}

\begin{figure}[ht!]
    \centering
    \includegraphics[width=0.9\linewidth]{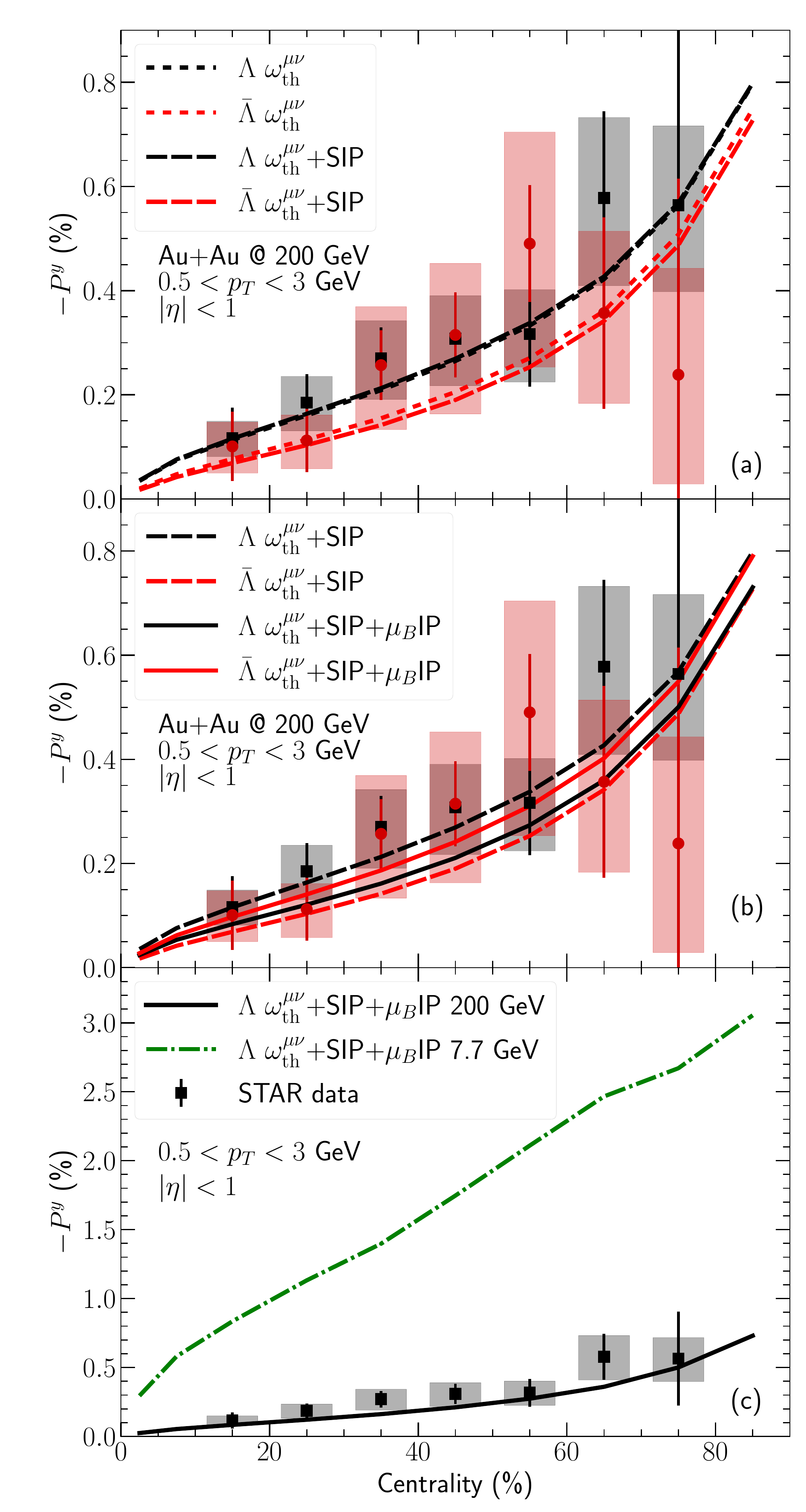}
    \caption{(Color online) Panels (a) and (b): The centrality dependence of the global $\Lambda$ polarization with different gradient terms in Au+Au collisions at 200 GeV compared with the STAR measurements \cite{Adam:2018ivw}. Panel (c): Model prediction for $P^y_\Lambda$ at 7.7 GeV. The STAR polarization data points are rescaled by 0.877 because the latest hyperon decay parameter $\alpha_\Lambda$ \cite{Zyla:2020zbs}.}
    \label{fig:Au200_Py_vs_cen}
\end{figure}
\begin{figure}[ht!]
    \centering
    \includegraphics[width=0.9\linewidth]{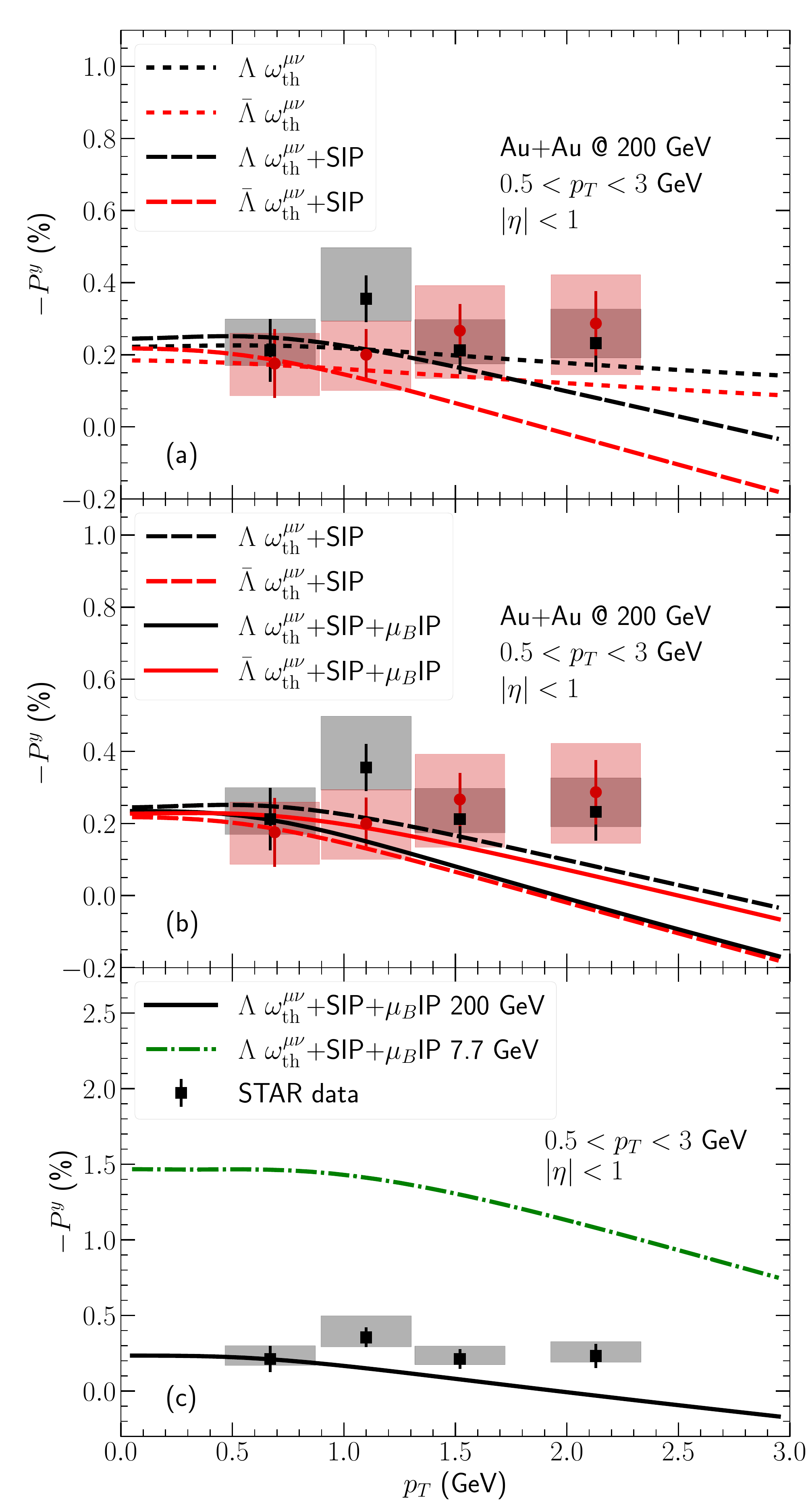}
    \caption{(Color online) Panels (a) and (b): The $p_T$-differential polarization for $\Lambda$ with different gradient terms in 20-60\% Au+Au collisions compared with the STAR measurements \cite{Adam:2018ivw}. Panel (c): Model prediction for $P^y_\Lambda(p_T)$ at 7.7 GeV. The STAR polarization data points are rescaled by 0.877 because the latest hyperon decay parameter $\alpha_\Lambda$ \cite{Zyla:2020zbs}.}
    \label{fig:Au200_Py_vs_pT}
\end{figure}

Figures~\ref{fig:Au200_yLfracDep} and \ref{fig:v1slope} show that the longitudinal rapidity fraction parameter $f$ can be tightly constrained by these two experimental observables.
Figure~\ref{fig:Py_vs_ecm} shows the main results of this work. We adjust the parameter $f$ at every collision energy to match the slope of the pion's directed flow at mid-rapidity and make predictions for $\Lambda$'s global polarization. We find that the $f$ increases from 0.15 to 0.5 as the collision energy goes down from 200 GeV to 7.7 GeV. A larger $f$ is needed at lower collision energy, indicating that more longitudinal momenta of the system are attributed to the initial longitudinal flow velocity at the lower collision energy. The initial density and velocity profiles for hydrodynamics are further away from the Bjorken boost-invariant assumption at the lower collision energy. With the parameter $f$ constrained by the pion's directed flow measurements, our model shows a reasonable description of the global polarization of $\Lambda$ and $\bar{\Lambda}$ in Fig.~\ref{fig:Py_vs_ecm}c.

With the constrained $f$ in our model, we can estimate the amount of OAM left in the fluid at mid-rapidity after the initial impact at different collision energies.
Based on OAM given by Eq. (\ref{eq:OAM_total_surface}),
Figure~\ref{fig:OAMfraction} shows that only about 0.5\% of the total OAM remains in the mid-rapidity region of 20-30\% Au+Au collisions at 200 GeV. This relative fraction of OAM increases as the collision energy goes down. At 7.7 GeV, the relative fraction increases up to $\sim15\%$ of the total OAM in the collision systems. Figure~\ref{fig:OAMfraction}b shows that although the total OAM increases with collision energy the absolute OAM in the mid-rapidity region remains around 100-200 $\hbar$ for 20-30\% Au+Au collisions from 7.7-200 GeV.

We make further comparisons with different gradient terms in the global polarization observables in Figs.~\ref{fig:Py_vs_ecm}c and \ref{fig:Py_vs_ecm}d. We note that thermal vorticity gives the dominant contribution to the global $\Lambda$ polarization. The shear-induced polarization is negligible, while the $\mu_B$-induced polarization flips the ordering between $\Lambda$ and $\bar{\Lambda}$'s polarization in all energies. This result demonstrates that the $\mu_B$ distribution inside fluid is important to determine the difference between the $\Lambda$ and $\bar{\Lambda}$'s polarization. This conclusion is inline with the finding in Ref.~\cite{Vitiuk:2019rfv}.

\begin{figure}[ht!]
    \centering
    \includegraphics[width=0.9\linewidth]{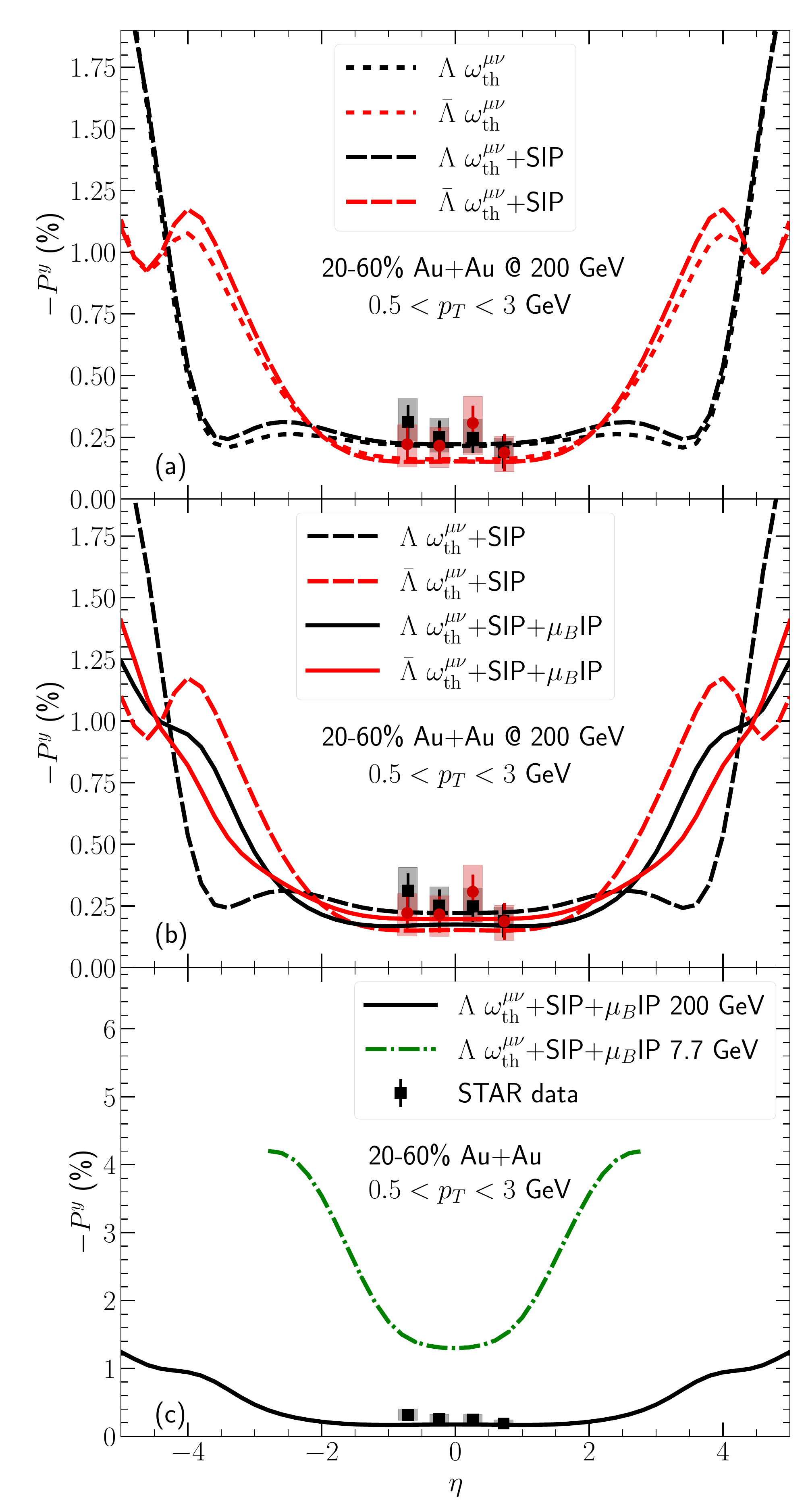}
    \caption{(Color online) Panels (a) and (b): The pseudorapidity dependence of the $\Lambda$ polarization with different gradient terms in 20-60\% Au+Au collisions at 200 GeV compared with the STAR measurements \cite{Adam:2018ivw}. Panel (c): Model prediction for $P^y_\Lambda(\eta)$ at 7.7 GeV. The STAR polarization data points are rescaled by 0.877 because the latest hyperon decay parameter $\alpha_\Lambda$ \cite{Zyla:2020zbs}.}
    \label{fig:Au200_Py_vs_eta}
\end{figure}

In Figs.~\ref{fig:Au200_Py_vs_cen}, \ref{fig:Au200_Py_vs_pT}, and \ref{fig:Au200_Py_vs_eta}, we further compare the centrality, $p_T$, and pseudorapidity dependence of $\Lambda$'s and $\bar{\Lambda}$'s global polarization with the STAR measurements at 200 GeV, respectively \cite{Adam:2018ivw}. 

Figures~\ref{fig:Au200_Py_vs_cen}a and \ref{fig:Au200_Py_vs_cen}b show that our model calculations provide a good description of the centrality dependence of the STAR data at 200 GeV. The $\mu_B$IP terms reverse the difference between $\Lambda$ and $\bar{\Lambda}$'s global polarization, which suggests that the evolution net baryon density and its gradients are crucial to understand the difference between $\Lambda$'s and $\bar{\Lambda}$'s global polarization. Figure~\ref{fig:Au200_Py_vs_cen}c further show our prediction for the $\Lambda$ polarization at 7.7 GeV with all the gradient terms included. 

In Figs.~\ref{fig:Au200_Py_vs_pT}a and \ref{fig:Au200_Py_vs_pT}b, we find that our results with only thermal vorticity has a weak $p_T$ dependence. According to Eq.~(\ref{eq:axialVector}), the SIP terms introduce a linear dependence of $P^y$ on hyperon's momentum. Because the total contribution from the SIP terms vanishes when integrating over the momentum, they enhance the $P^y$ in small $p_T$ but suppress it for $p_T > 1$ GeV. Despite the current STAR measurements contains significant uncertainties, our results with SIP show a stronger $p_T$ dependence than the data. In the meantime, the $\mu_B$IP terms invert the ordering between $\Lambda$ and $\bar{\Lambda}$. Figure \ref{fig:Au200_Py_vs_pT}c shows our prediction at 7.7 GeV which has the same $p_T$ dependence as those in the 200 GeV.

Figures~\ref{fig:Au200_Py_vs_eta}a and \ref{fig:Au200_Py_vs_eta}b show the pseudo-rapidity distribution of the global polarization for $\Lambda$ and $\bar{\Lambda}$ at 200 GeV with different gradient terms. Both $P^y_\Lambda$ and $P^y_{\bar{\Lambda}}$ have a plateau structure within $\vert \eta \vert < 2$. Using thermal vorticity results in a slightly larger polarization for $\Lambda$ than that of $\bar{\Lambda}$. In the forward and backward rapidity regions $\vert \eta \vert > 2$, the magnitudes of $P^y$ increase rapidly in our model. The $\mu_B$IP terms give different contributions to $\Lambda$ and $\bar{\Lambda}$ and reduce the difference in the forward and backward rapidity regions.

We further provide our model prediction with all the gradient terms included for 7.7 GeV in Fig.~\ref{fig:Au200_Py_vs_eta}c. The plateau window of $\Lambda$'s polarization shrinks as the collision energy goes down. At 7.7 GeV, the $P^y_\Lambda$ remains approximately constant within $\vert \eta \vert < 1$ and increases in the forward and backward rapidity regions.

\section{Conclusions} \label{sec:conc}

In this work, we develop a hybrid dynamical framework, which explicitly conserves energy, momentum, and orbital angular momentum from the initial collision geometry to the following hydrodynamic evolution. We introduce the longitudinal rapidity fraction parameter $f$ to vary how local net longitudinal momentum is distributed to flow velocity and energy density rapidity profile. This model parameter controls the amount of fluid vorticity correlated with the initial OAM at the beginning of the hydrodynamics. We study the evolution of the fluid vorticity during the hydrodynamic phase and find that the fluid expansion monotonically reduces the space-time averaged fluid vorticity as a function of time. Therefore, the initial distribution of fluid vorticity has a strongly correlation with their values at particlization and the magnitude of the hyperon's global spin polarization.

Our phenomenological studies have shown that the pion's directed flow and global polarization of $\Lambda$ hyperons together can set strong constraints on the size of initial longitudinal flow velocity at different collision energies. By fitting the STAR measurements, we quantify the amount of orbital angular momentum left in the midrapidity fluid after the initial impact. We find that about 0.5\% of the total OAM remains at the mid-rapidity for 20-30\% Au+Au collisions at 200 GeV, and this relative fraction increases to $\sim$15\% at 7.7 GeV.  
The centrality, $p_T$, and pseudorapidity dependence of $P^y_\Lambda$ show reasonable agreement with the STAR measurements at 200 GeV.

We further quantify the effects of new gradient terms proposed in Refs.~\cite{Hidaka:2017auj, Liu:2020dxg, Liu:2021uhn, Becattini:2021suc} on the global spin polarization of $\Lambda$ hyperons. The global polarization $P^y_\Lambda$ receives the dominant contribution from the fluid's thermal vorticity at the particlization hyper-surface.
The shear-induced polarization introduces a sizable $p_T$ dependence to $\Lambda$'s global polarization, while its net effect on the integrated polarization is small. The $\mu_B$-induced polarization can alter the ordering between $\Lambda$ and $\bar{\Lambda}$'s global polarization, which indicates that the difference between $\Lambda$ and $\bar{\Lambda}$'s global polarization may not be related to a non-zero magnetic field at freeze-out. A similar conclusion is made in Ref.~\cite{Vitiuk:2019rfv}.

\section*{Acknowledgments}
We thank Sean Gavin, Cheming Ko, Michael Lisa, George Moschelli, Jun Takahashi, Giorgio Torrieri, Sergei Voloshin, and Yi Yin for fruitful discussion. This work is supported in part by the U.S. Department of Energy (DOE) under grant number DE-SC0013460 and in part by the National Science Foundation (NSF) under grant number PHY-2012922.
This research used resources of the National Energy Research Scientific Computing Center, which is supported by the Office of Science of the U.S. Department of Energy under Contract No. DE-AC02-05CH11231, resources provided by the Open Science Grid, which is supported by the National Science Foundation and the U.S. Department of Energy's Office of Science, and resources of the high performance computing services at Wayne State University.
This work also is supported by the U.S. Department of Energy, Office of Science, Office of Nuclear Physics, within the framework of the Beam Energy Scan Theory (BEST) Topical Collaboration.

\appendix
\section{Estimate spin polarization with different vorticity tensors}
\label{sec:appendixA}

\begin{figure}[ht!]
    \centering
    \includegraphics[width=0.9\linewidth]{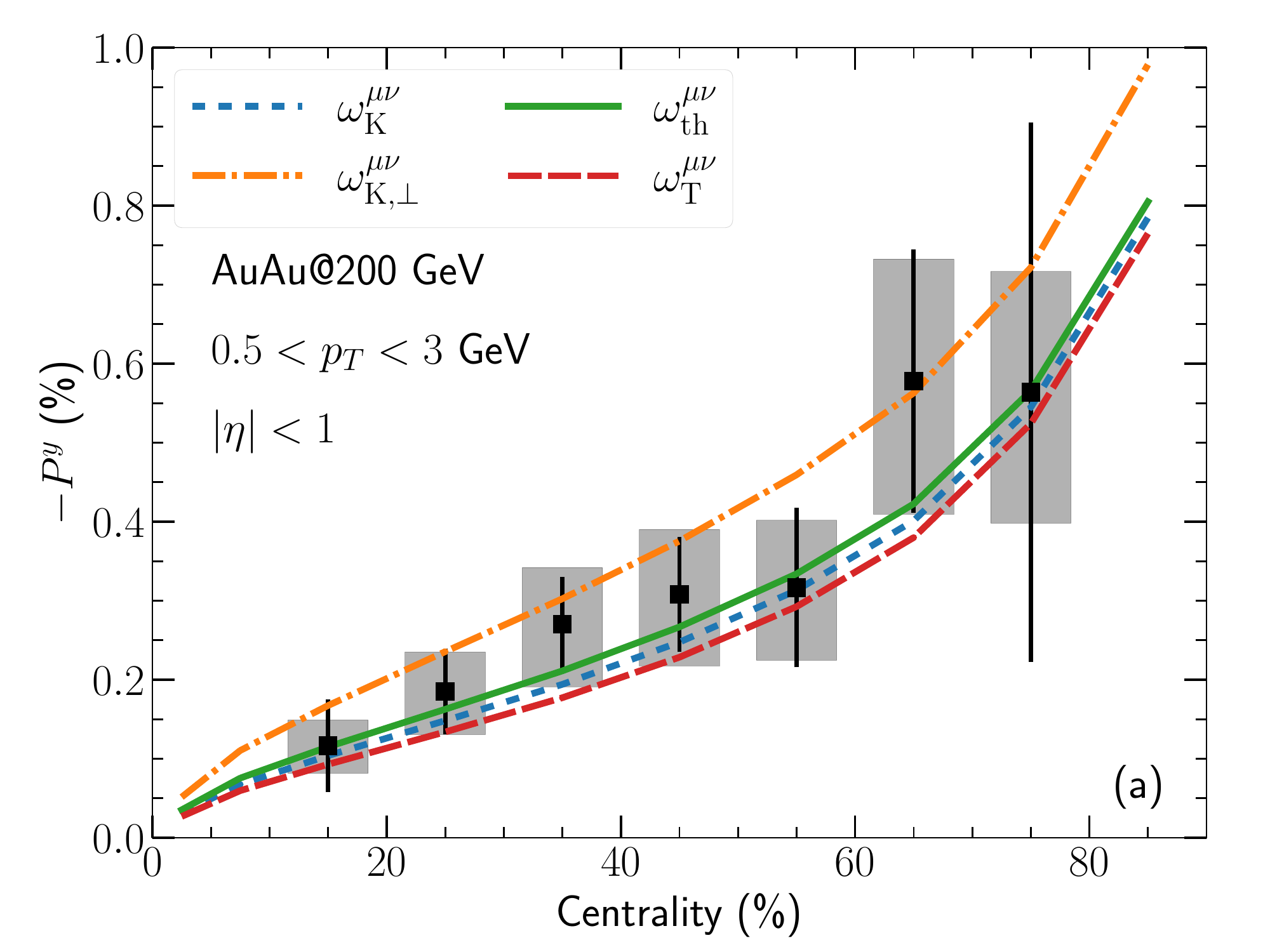}
    \includegraphics[width=0.9\linewidth]{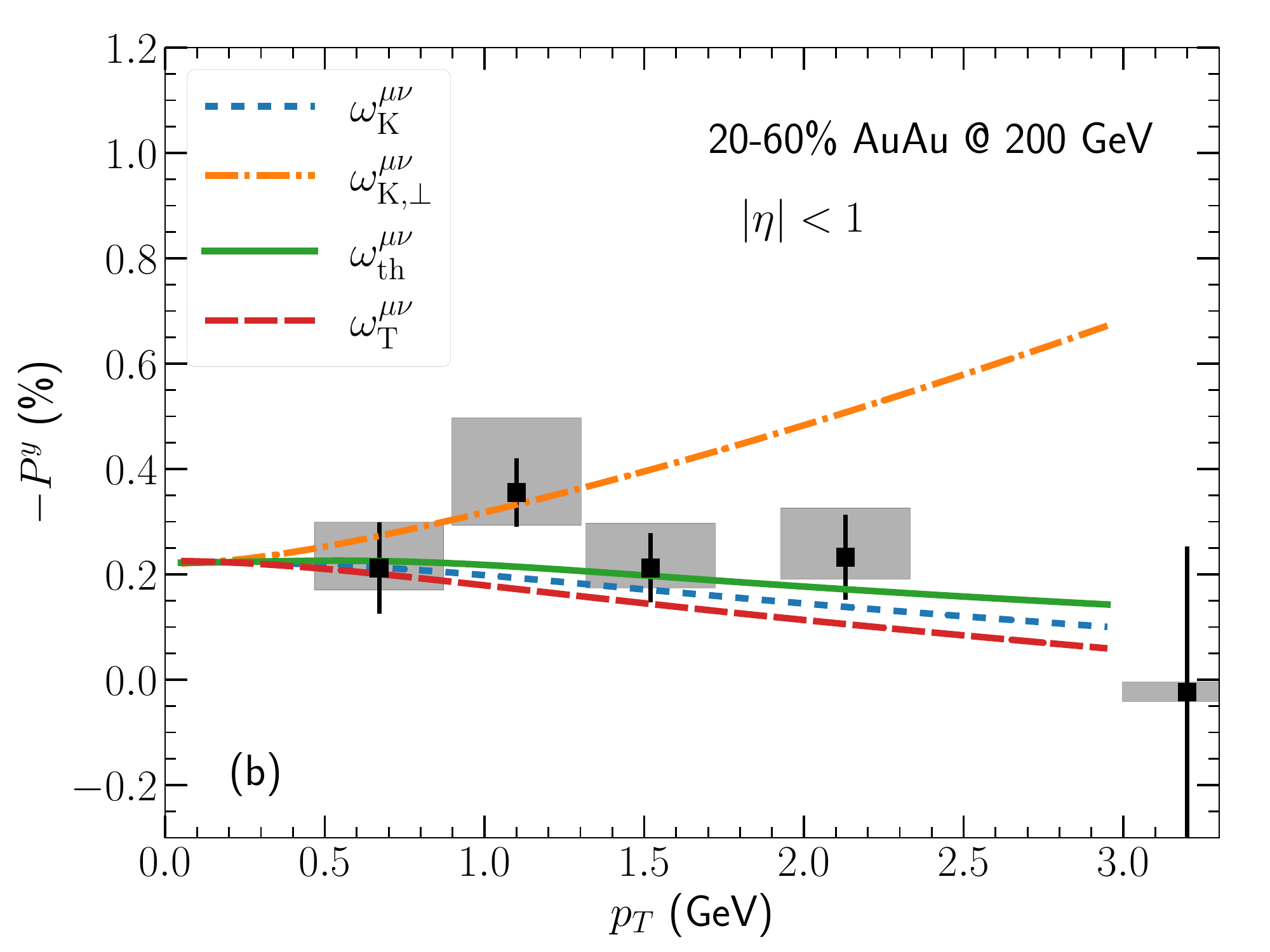}
    \includegraphics[width=0.9\linewidth]{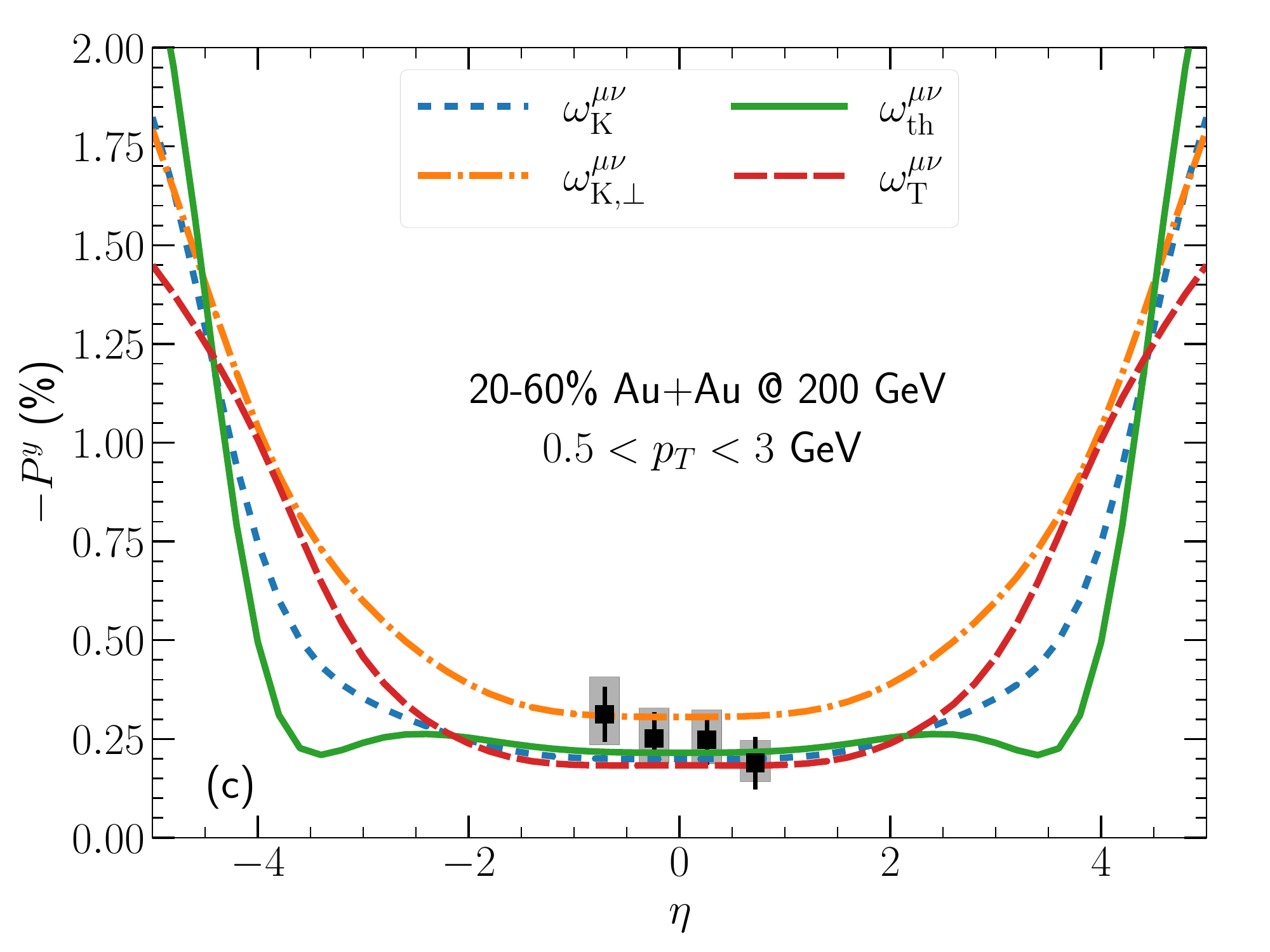}
    \caption{(Color online) The global $\Lambda$ polarization computed with different vorticity tensors with $f = 0.15$ in Au+Au collisions at 200 GeV compared with the STAR measurements \cite{Adam:2018ivw}. The STAR polarization data points are rescaled by 0.877 because the latest hyperon decay parameter $\alpha_\Lambda$ \cite{Zyla:2020zbs}.}
    \label{fig:Au200_vorDep}
\end{figure}

Within fluid dynamical evolution, different types of vorticity tensors can be defined as those in Eqs.~(\ref{eq:omega_K})-(\ref{eq:omega_T}). The authors in \cite{Wu:2019eyi} proposed that calculating $\Lambda$ spin polarization with the $T$-vorticity could reproduce the correct azimuthal dependence of the longitudinal polarization measured by the STAR Collaboration \cite{Adam:2019srw}. It is possible that the hyperon's spin polarization could be related to these fluid vorticity tensors. In this appendix, we will compute the $\Lambda$'s global polarization with the vorticity tensors defined Eqs.~(\ref{eq:omega_K})-(\ref{eq:omega_T}),
\begin{eqnarray}
    S^\mu(p^\mu) = -\frac{1}{8m} \frac{\int d^3 \Sigma_\alpha p^\alpha n_0(E) (1 - n_0(E)) \epsilon^{\mu\nu\alpha\gamma} p_\nu \Omega_{\alpha \gamma}}{\int d^3 \Sigma_\alpha p^\alpha n_0(E)}, \nonumber \\
    \label{eq:SpinVector_vor}
\end{eqnarray}
where $\Omega^{\alpha \gamma} = \frac{\omega^{\alpha \gamma}_{K}}{T}, \frac{\omega^{\alpha \gamma}_{K, \perp}}{T}, \omega^{\alpha \gamma}_\mathrm{th}, \frac{\omega^{\alpha \gamma}_{T}}{T^2}$ \cite{Wu:2019eyi}. We interpret their relative variations as the theoretical uncertainties in our calculations. The SIP's and $\mu_B$IP's contributions remains the same as those shown in Figs.~\ref{fig:Au200_Py_vs_cen}-\ref{fig:Au200_Py_vs_eta}.

Figure~\ref{fig:Au200_vorDep} shows the centrality, $p_T$, and pseudorapidity dependence of global $\Lambda$ polarization computed with different types of vorticity tensor. The kinematic, thermal, and $T$ vorticity tensors give very close results of $P^y_\Lambda$ as functions of centrality, $p_T$, and pseudorapidity within $\vert \eta \vert < 2$. These results means that the temperature gradients do not generate a significant contribution to the azimuthally integrated global polarization. The transverse kinematic vorticity differs from the kinematic vorticity by the fluid acceleration, as shown in Eq.~(\ref{eq:omega_KSP}). The difference between the results from these two vorticity tensors shows that the fluid acceleration suppresses the overall magnitude of global polarization by $\sim 40\%$. This suppression grows with $p_T$ as shown in Fig.~\ref{fig:Au200_vorDep}b.

\section{The freeze-out energy density dependence on $\Lambda$'s global polarization}
\label{sec:appendixB}

\begin{figure}[ht!]
    \centering
    \includegraphics[width=0.9\linewidth]{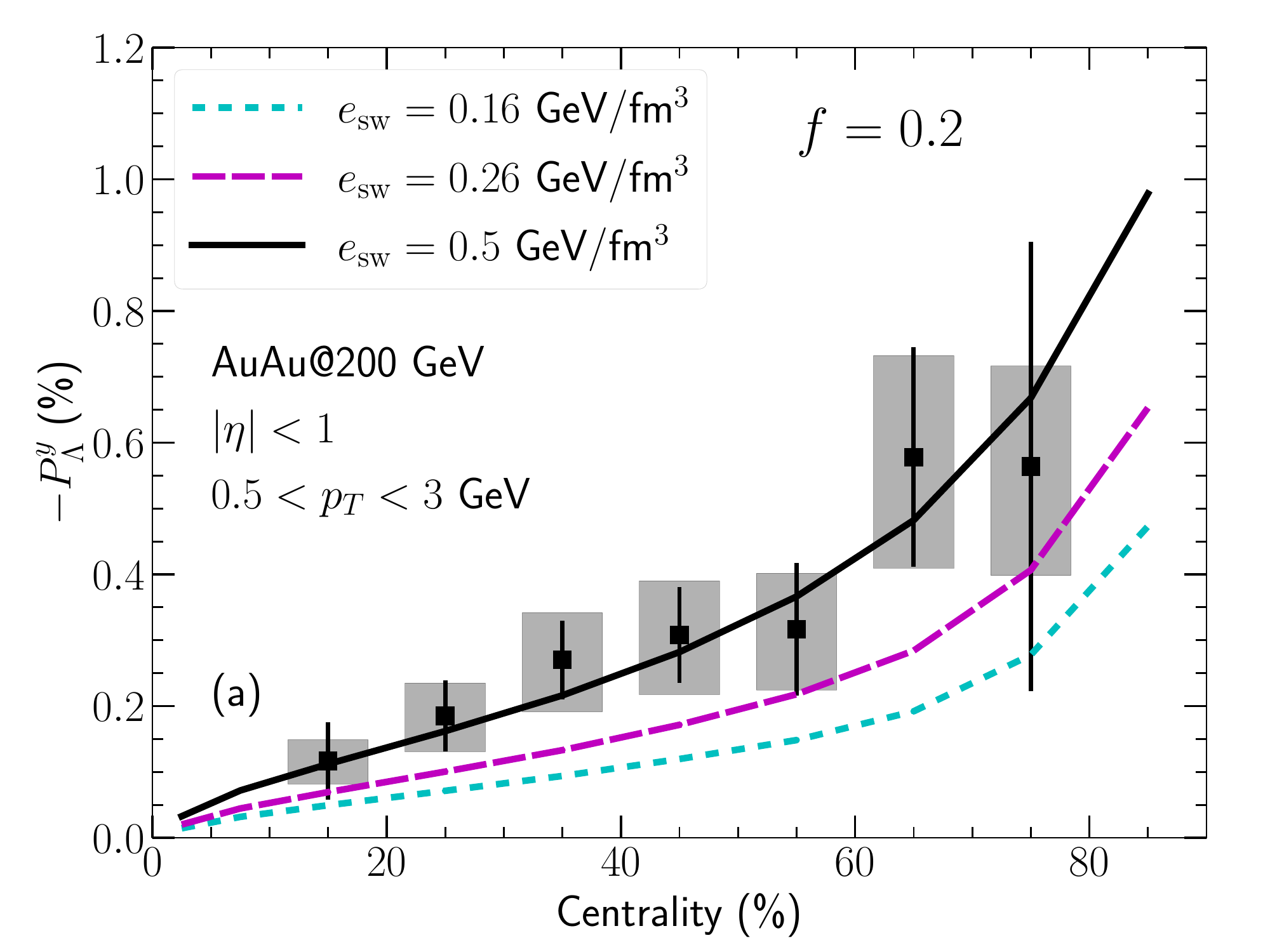}
    \includegraphics[width=0.9\linewidth]{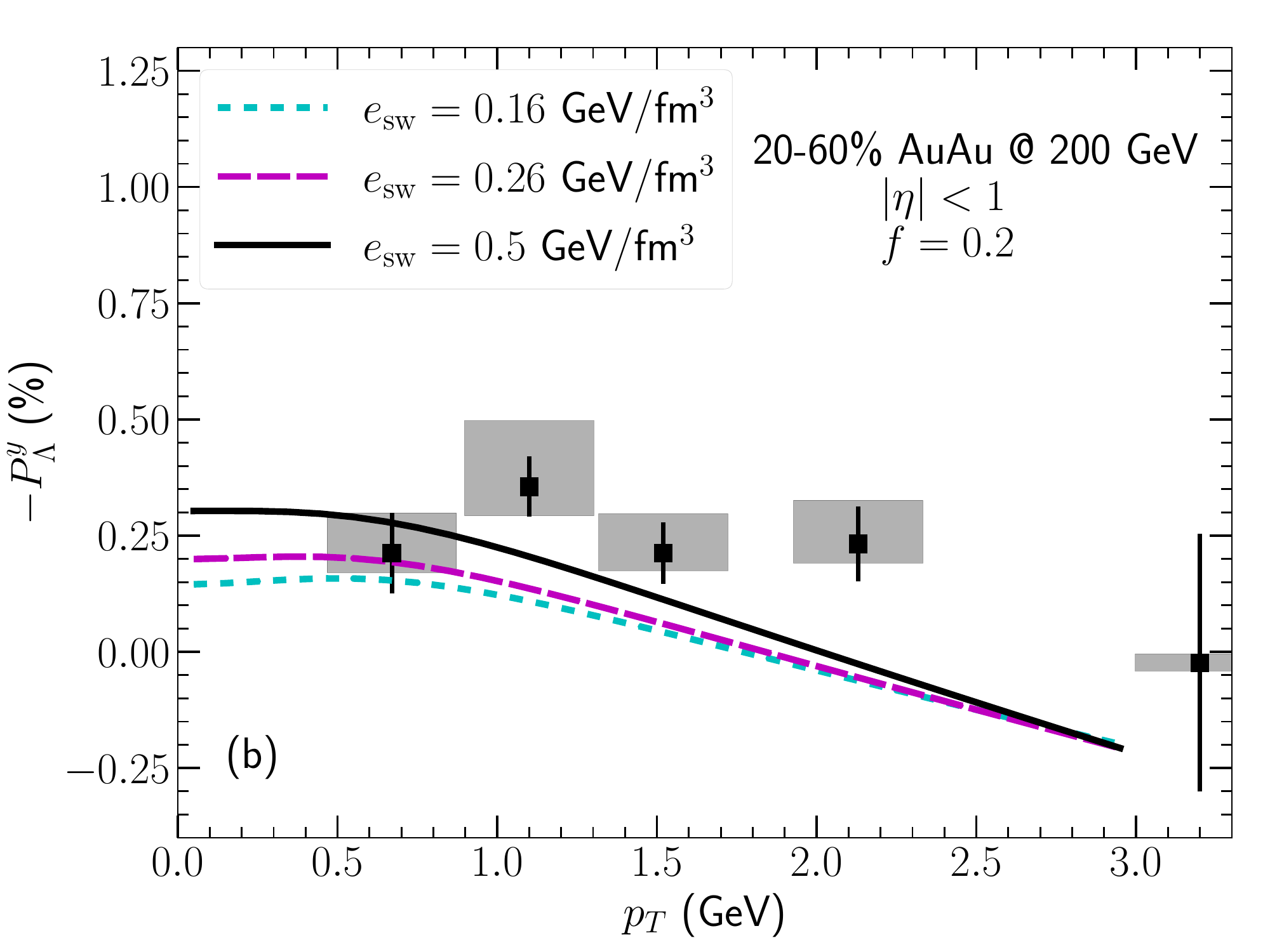}
    \includegraphics[width=0.9\linewidth]{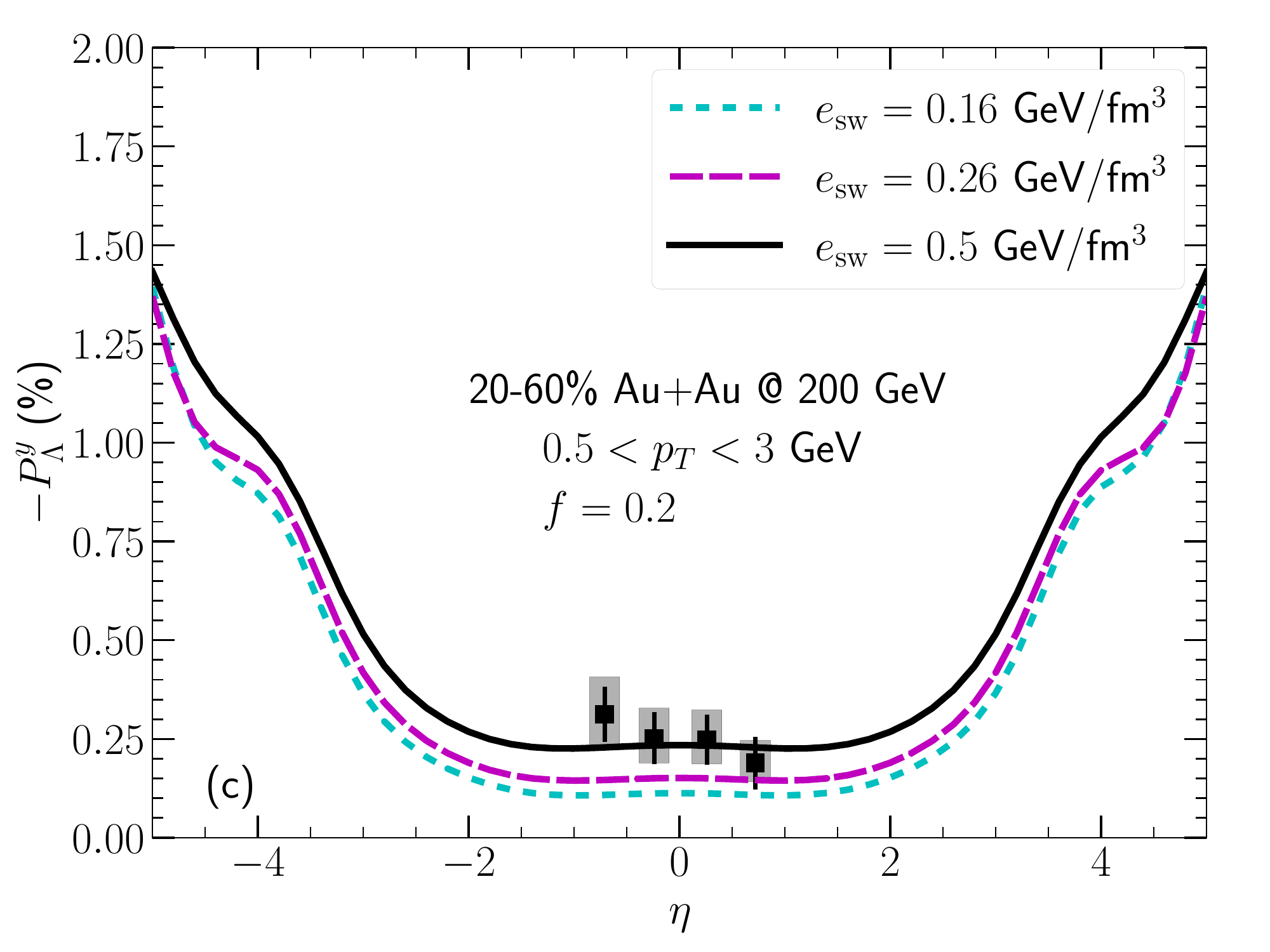}
    \caption{(Color online) The global $\Lambda$ polarization's dependence on the switching energy density in Au+Au collisions at 200 GeV compared with the STAR measurements \cite{Adam:2018ivw}. The STAR polarization data points are rescaled by 0.877 because the latest hyperon decay parameter $\alpha_\Lambda$ \cite{Zyla:2020zbs}.}
    \label{fig:Au200_eswDep}
\end{figure}

In hydrodynamic + hadronic transport models, the spin polarizations of $\Lambda$ and $\bar{\Lambda}$ hyperons are often computed at the particlization hypersurface but not at kinetic freeze-out because it is difficult to track and model the spin information in the microscopic hadronic transport models. In this appendix, we explore the sensitivity of the $\Lambda$'s global polarization on the particlization energy density of hypersurface.

Figure~\ref{fig:Au200_eswDep} shows how the global $\Lambda$ polarization depends on the switching energy density. The overall magnitudes of the global polarization of $\Lambda$ decrease with the $e_\mathrm{sw}$, which is the consequence of smaller fluid gradients on the switching hypersurface with lower $e_\mathrm{sw}$. The gradients of temperature and flow velocity decrease roughly as $1/\tau$ at late time of the hydrodynamic evolution \cite{Vujanovic:2019yih}. Because the fireball lives longer with a lower switching energy density, the magnitudes of thermal vorticity tensors decreases with $e_\mathrm{sw}$ as indicated in Fig.~\ref{fig:vor_evo}.

Figure~\ref{fig:Au200_eswDep}a shows that the $P^y_\Lambda$ as function of centrality is 5-10\% smaller with the smaller $e_\mathrm{sw}$.
In addition to the overall suppression, the shape of $P^y_\Lambda(p_T)$ gets flatter at lower switching energy density as shown in Fig.~\ref{fig:Au200_eswDep}b. The change in the $p_T$ dependence is caused by a larger radial flow as the fireball evolves longer to the lower $e_\mathrm{sw}$ hypersurface. The stronger radial flow blue-shifts more $\Lambda$ to high $p_T$, flattening the $P^y_\Lambda(p_T)$. Finally, Figure~\ref{fig:Au200_eswDep}c shows that a lower $e_\mathrm{sw}$ hypersurface results in an overall suppression of   $P^y_\Lambda(\eta)$ with the $\eta$-dependence roughly unchanged.

\bibliography{vorticity_BES}

\end{document}